\documentclass[final,authoryear, times, 3p]{elsarticle}

\makeatletter
\journal{Journal of Economic Dynamics and Control}
\def\ps@pprintTitle{%
	\let\@oddhead\@empty
	\let\@evenhead\@empty
	\def\@oddfoot{}%
	\let\@evenfoot\@oddfoot}
\makeatother

\usepackage{numcompress}\biboptions{authoryear}

\usepackage{epstopdf}
\usepackage{graphicx} 
\usepackage{amsfonts}
\usepackage{amsmath} 
\usepackage{bm}
\usepackage{amssymb} 
\usepackage{siunitx} 
\usepackage{booktabs} 
\usepackage{cleveref}
\usepackage{float}
\usepackage[caption = false]{subfig} 
\usepackage{multirow}
\usepackage{booktabs}
\usepackage{tabu}
\usepackage{color}
\usepackage{todonotes}
\usepackage{lmodern}

\usepackage{mathtools}
\DeclarePairedDelimiter\ceil{\lceil}{\rceil}

\newdefinition{defi}{Definition}

\usepackage{color} 

\begin{document}

\begin{frontmatter}
\title{What is the Minimal Systemic Risk in Financial Exposure Networks?}

\author[wu]{Christian Diem} \ead{christian.diem@s.wu.ac.at}
\author[inet,math,csh]{Anton Pichler} \ead{anton.pichler@wolfson.ox.ac.uk} 
\author[cosy,csh,sfi,iiasa]{Stefan Thurner\corref{cor}} \ead{stefan.thurner@muv.ac.at} 

\cortext[cor]{Corresponding author}

\address[wu]{Institute for Statistics and Mathematics, Vienna University of Economics and Business, Welthandelsplatz 1, A-1020, Austria}
\address[csh]{Complexity Science Hub Vienna, Josefst\"adter Stra\ss e 39, A-1080, Austria}
\address[inet]{Institute for New Economic Thinking, University of Oxford, Manor Road, OX1 3UQ, UK}
\address[math]{Mathematical Institute, University of Oxford, Woodstock Road, Oxford OX1 3LP, UK}
\address[iiasa]{IIASA, Schlossplatz 1, A-2361 Laxenburg, Austria}
\address[cosy]{Section for Science of Complex Systems, Medical University of Vienna, Spitalgasse 23, A-1090, Austria} 
\address[sfi]{Santa Fe Institute, 1399 Hyde Park Road, Santa Fe, NM 87501, USA} 
	
\begin{abstract}
Management of systemic risk in financial markets is traditionally associated with setting (higher) capital requirements for market participants. There are indications that while equity ratios have been increased massively since the financial crisis, systemic risk levels might not have lowered, but even increased (see ECB data\footnote{ECB Statistical Data Warehouse: Consolidated banking data set}; SRISK time series\footnote{https://vlab.stern.nyu.edu/welcome/risk/}). It has been shown that systemic risk is to a large extent related to the underlying network topology of financial exposures. A natural question arising is how much systemic risk can be eliminated by optimally rearranging these networks and without increasing capital requirements. Overlapping portfolios with minimized systemic risk which provide the same market functionality as empirical ones have been studied by \cite{pichler2018systemic}. Here we propose a similar method for direct exposure networks, and apply it to cross-sectional interbank loan networks, consisting of 10 quarterly observations of the Austrian interbank market. We show that the suggested framework rearranges the network topology, such that systemic risk is reduced by a factor of approximately 3.5, and leaves the relevant economic features of the optimized network and its agents unchanged. The presented optimization procedure is not intended to actually re-configure interbank markets, but to demonstrate the huge potential for systemic risk management through rearranging exposure networks, in contrast to increasing capital requirements that were shown to have only marginal effects on systemic risk \cite[]{poledna2017basel}. 
Ways to actually incentivize a self-organized formation toward optimal network configurations were introduced in \cite{thurner2013debtrank} and \cite{poledna2016elimination}.
 For regulatory policies concerning financial market stability the knowledge of minimal systemic risk for a given economic environment can serve as a benchmark for monitoring actual systemic risk in markets. 

\end{abstract}


\end{frontmatter}

\section{Introduction}
Increasing capital requirements for market participants is an obvious suggestion for improving the resilience of financial systems, and in particular for  reducing systemic risk in financial markets.
Examples for such policies, where capital requirements depend on macro prudential regulation are \cite{cont2010network}, who propose capital requirements in relation to the Contagion Index values of banks,  \cite{gauthier2012macroprudential} suggest that bank capital buffers should correspond to their contributions to overall systemic risk, or \cite{markose2012systemic} proposes a capital surcharge related to the eigenvector centrality of banks in the financial network. Also in the classical risk measure literature following \cite{artzner1999coherent} and \cite{follmer2002convex} the risk of an asset is measured by the amount of capital that needs to be added to the position in order to make the position acceptable to the regulator, or to the firm itself. This approach can be extended to determine the capital requirements for financial institutions to bring systemic risk to levels, which are acceptable to the regulator, see e.g. \cite{feinstein2017measures} or \cite{biagini2015unified}. 

In the recent past, after the last financial crisis, bank capital requirements have been adjusted upwards. In the Basel III accord the regulatory minimum capital requirements for Common Equity Tier 1 (CET1) have been increased from 2\% to 4.5\%, and Tier 1 Capital from 4\% to 6\% 
\cite{basel3reg}. Additionally, a capital conservation buffer has been introduced by increasing CET1 and Tier 1 capital further to 7\% and 8.5\%, respectively. On top of this, national authorities can set an additional counter-cyclical buffer in the range between zero and 2.5\%  for phases of excessive credit growth. Global systemically important institutions have to meet additional CET1 requirements in a range of 1\% to 2.5\% \cite{basel3gsib}. 

Bank capital levels have been steadily increasing since the introduction of these new regulations. The monitoring report of the Basel Committee \cite{basel3monitoring} shows that for a sample of 86 international banks with $\text{Tier 1 Capital larger than}  \; \$3\text{bn}$ the CET1 increased from 7.2\% to 12.7\% in the period from 2011 to 2018 \cite[Graph 15]{basel3monitoring}. For Germany, Spain, France, and Italy ECB data shows increases of Tier 1 capital ratios  from 9.2\%, 8.1\%, 8.4\% and 6.9\% to 16.4\%, 13.2\%, 15.3\%, and 14.4\%, respectively for the period from 2008 to 2017\footnote{ECB Statistical Data Warehouse: Consolidated banking data set}. Nonetheless, some indicators of systemic risk suggest that systemic risk levels are not declining, but are still substantially higher than before the financial crises. A prominent example is the SRISK indicator of \cite{brownlees2016srisk}, which shows that the systemic risk level in Europe is twice as high than before the crises\footnote{SRISK levels for different regions are provided by https://vlab.stern.nyu.edu/welcome/risk/}. 

However, capital levels for absorbing shocks are only one part of the story in the context of systemic risk. The other essential component that determines systemic risk is the exposure network that is generated by contracts between financial agents. In particular, these networks capture the risks of potential cascading events that could threaten large fractions of financial markets to fail. This fact is reflected in a number of works such as in \cite{allen2000financial}, \cite{freixas1998systemic}, \cite{eisenberg2001systemic},  \cite{boss2004network}, \cite{cont2010network}, \cite{gai2010contagion}, \cite{battiston2012debtrank}, \cite{markose2012to} \cite{thurner2013debtrank},  \cite{glasserman2015likely}. 

It is therefore natural to ask what contributions to systemic risk originate specifically from networks, and how their topology influences systemic risk. Indeed, many contributions to the systemic risk literature investigate the effect of network characteristics on systemic risk. \cite{allen2000financial} compare the effects of different network topologies, such as rings, fully connected graphs, and interconnected subgroups on the interbank market stability.   In \cite{boss2004network} the role of scale free network topologies in the context of systemic risk and stability is discussed. In  \cite{boss2004contagion}  the  betweenness centrality measure is introduced as a network-based measure for systemic risk. \cite{nier2007network} investigate the effects of network connectivity and concentration  on contagious defaults. \cite{gai2010contagion} employ a stylized analytical contagion model and look at the fraction of defaulting banks for given average degrees. \cite{puhr2012contagiousness} employ panel regressions to study the effects of network measures like Katz centrality on the number of defaulting banks, which are obtained from a simulation study. The concept of \textit{too interconnected to fail} is also part of this discussion and is investigated for example by \cite{markose2012to}. \cite{glasserman2016contagion} dedicate a considerable part of their literature review to this topic. These and many more theoretical and empirical works indicate the possibility to use networks of financial connections as a leverage point for reducing systemic risk in a financial system as an effective alternative to costly capital requirements that were shown to have limited effects on systemic risk reduction \cite[]{poledna2017basel}. If systemic risk can indeed be effectively reduced by altering the underlying exposure network characteristics, this should be prominently factored into financial market stability policies. It is therefore essential to systematically estimate the full potential for network-based systemic risk reduction.

In this work we propose a method for quantifying the systemic risk reduction potential in empirically observed direct exposure networks by employing standard optimization techniques. The systemic risk of a network is measured with the so-called DebtRank \cite[]{battiston2012debtrank}. The actual optimization relies on an approximation of the DebtRank, which is computed iteratively and is thus hard to be used in optimization. The approximation is based on the direct impacts of defaulting banks on their neighboring nodes in the exposure network. We show how the systemic risk optimization can be solved as a mixed integer linear program (MILP) by standard reformulation techniques. The optimization problem can be solved by state of the art optimization algorithms and could therefore also be easily implemented in practice. In the empirical part of this study we show the effectiveness of the proposed method by applying it to a  data set containing ten quarterly observations of the Austrian interbank liability network  from 2006 to 2008. Our findings for the largest 70 banks suggest that the DebtRank of individual banks can be reduced on average by a factor of 3.5. This means sizeable reductions of the DebtRank for almost all of the 70 banks across the ten quarters with only few exceptions. 

In practice, due to the current lack of incentive schemes for systemic risk management \cite[]{leduc2017incentivizing}, financial networks do not evolve toward systemically optimal configurations, and obviously they do not result in any way from such optimization procedures. However, our study can give an estimate for the systemic risk reduction potential stemming from a specific reorganization of empirically observed networks. The same optimization algorithm can be used to compute network configurations that yield a maximum of overall systemic risk. In this way, for any observed financial network, the proposed optimization procedure yields a ``range'' of network structures, corresponding to minimal and maximum DebtRank. This allows us to identify network characteristics that are typical for low, medium, and high DebtRank. 
 
Closely related studies include \cite{poledna2016elimination} and \cite{leduc2017incentivizing}, which investigate how systemic risk can be reduced by changing the underlying networks, when financial agents are incentivized to favor transactions with low systemic risk  in the network. The idea of applying network optimization techniques that are a commonly used in operations research to systemic risk reduction is relatively new. It has been pioneered in the specific context of overlapping portfolios and fire sales by \cite{pichler2018systemic}
who find reductions of systemic risk of around 50\% by rearranging the network structure of the European government bond market. The optimization approach there however -- a quadratically constrained quadratic program (QCQP) -- is substantially different from the one presented here.  A recent paper by  \cite{krause2019} focuses on  small homogeneous macroeconomic shocks affecting the assets of all banks simultaneously and how these shocks are amplified in the banking system.  They show a Monte Carlo algorithm for finding minimal and maximal networks with respect to the amplification of such small homogeneous macro shocks. Another related study is \cite{aldasoro2017bank}. The authors employ a theoretical model of the interbank network where optimizing risk-averse banks invest in illiquid assets and lend to each other. In their model they account for contagion originating from liquidity hoarding, interbank interlinkages and fire sales. Their model leads to a specific interbank network for which properties of the network topology are reported. 

The paper is organized as follows. Section \ref{quantifyingSR} presents our approach to quantify systemic risk. In Section \ref{method} we derive the optimization problem for reducing DebtRank. We discuss the data and the results of the application to the Austrian interbank market in detail in Section \ref{data}, before we conclude in Section \ref{conclusion}.

\section{Quantifying systemic risk} \label{quantifyingSR}

Quantification of systemic risk in financial networks is a non-trivial task and depends on the specific aspects of interest. Based on very different ideas, various systemic risk measures were suggested. Some such as those that are based on networks, were already mentioned above. 
Other well known approaches include the $\Delta$CoVaR, that measures the tail dependence of bank asset returns by \cite{adrian2016covar}, systemic expected shortfall (SES) by \cite{acharya2017measuring}  measuring the tendency of a bank being undercapitalized if the whole system is undercapitalized,  the SRISK measure proposed by \cite{brownlees2016srisk}, or the put option portfolio approach by \cite{lehar2005measuring}. The advantage of these market based measures for systemic risk measurement is that they do not require the detailed (often restricted) information of financial networks but estimate systemic risk from openly accessible data. These models lack the possibility to estimate the contributions from cascading effects through  financial exposure networks. The difference between these two strands of literature is emphasized by \cite{benoit2017risks}.  

Here we choose the network-based measure DebtRank as a way to quantify systemic risk. The following method for minimizing systemic risk is then applicable to all direct financial exposure networks, whenever DebtRank is used as the measure for systemic risk. Examples for analyzing systemic risk on networks include interbank networks \cite[]{battiston2012debtrank, thurner2013debtrank}, derivatives and foreign exchange \cite[]{poledna2015the}, and credit-default swaps \cite[]{leduc2017systemic}. Without loss of generality for any kind of direct exposure network, we demonstrate the method for interbank asset-liability networks. 

We model the interbank market with $N$ banks as a directed weighted network represented by the asset-liability matrix, $L$. The nodes represent banks, links are the liabilities between banks. If bank $j$ lends $L_{ij}$ (monetary units) to bank $i$, we represent this as a directed link from  node $i$ to node $j$ with a corresponding weight of $L_{ij}$. $L_{ij}$ is $j$'s exposure towards $i$, i.e. if $i$ defaults the amount $L_{ij}$ is at risk for $j$. We denote the total interbank liabilities of bank $i$ to all others in the network  by $l_i = \sum_{j=1}^{N} L_{ij}$; the sum of all loans from $i$ to other banks is $a_i= \sum_{j=1}^{N} L_{ji}$. The equity of bank $i$ is denoted by $e_i$, and the total interbank market volume in the network is $\bar{L} = \sum_{i=1}^{N} l_i = \sum_{i=1}^{N} a_i$. The relative weight of bank $i$ in the network is $v_i = {a_i}/\bar{L}$.

In the case of the default of $i$, we assume that bank $j$ needs to write off $L_{ij}$ of its assets\footnote{For simplicity we assume zero recovery. Note that this assumption is not entirely unrealistic for short time scales, and is frequently used in the literature.}. Since a bank cannot have negative equity, the maximum impact that $i$ can have on $j$ is $e_j$. This motivates the definition of the direct impact matrix,
\begin{equation} \label{impactmat}
 W_{ij} = \min \left( \frac{L_{ij}}{e_j}, 1 \right) \quad, 
\end{equation}
which denotes the share of $j$'s equity lost due to the default of bank $i$. 

As stated above, we quantify systemic risk by using DebtRank. DebtRank is a recursive centrality measure designed specifically for networks of direct financial exposures and quantifies the impact of bank $i$ on the entire network if $i$ defaults. Every bank $i$ has a DebtRank value, $R_i$, between zero and one; $R_i=0$ means that bank $i$ has no impact on other banks, whereas $R_i=1$ indicates that the entire interbank asset weighted equity of the system is at risk, should $i$ default\footnote{From the definition of DebtRank it becomes obvious that $R_i=1$ can only occur if the weight $v_i = 0$. Thus, in most cases $R_i$ is strictly smaller than one.}. In that sense, $R_i$ is the fraction of the affected total value in the network by $i$'s default.

\begin{defi}[DebtRank]
DebtRank is defined by an iterative procedure that involves two state variables, $h$ and $s$. $h_i(t)$ measures the level of distress at iteration $t$; it is the fraction of equity, $e_i$, that was lost due to the default of other banks before $t$. Consequently, $h_i(t) \in [0,1]$, where $h_i(t)=1$ means default. The variable $s_j(t) \in \{U,D,I\}$ takes one of three states: undistressed, distressed, and inactive. The variables are initialized for $t=1$, as $h_i(1)= \delta_{ij}$,\footnote{$\delta_{ij}$ is the Kronecker symbol, $\delta_{ij}=1$ if $i=j$, and $\delta_{ij}=0$, otherwise.},  where $j$ is the bank, which initially defaults, and $s_i(1) = D$, for $i=j$ and $s_i(1) = U$, for $i\neq j$. The dynamics for the two state variables for $t \geq 2$ is defined by first updating $h_i(t)$ simultaneously for all $i$, followed by an update of $s_i(t)$, for all $i$. The update rules are given by
\begin{equation}
h_i(t) = \min \left( 1, h_i (t-1) + \sum_{j} W_{ji} h_j (t-1) \right)  
\quad , 
\end{equation}
where the summation over $j$ runs over all $j$, for which $s_j(t-1)=D$, and
\begin{equation}
s_i(t) = \begin{cases}
D \qquad \quad \; \; \text{if} \; h_i(t)>0 \; ; \; s_i(t-1) \neq I \\
I \qquad \qquad \text{if} \; s_i(t-1) = D \\
s_i(t-1) \quad \text{otherwise}  \quad .
\end{cases}
\end{equation}
The iterative procedure ends after $T$ steps at which all nodes are either undistressed or inactive. The DebtRank of bank $i$ is defined as 
\begin{equation} \label{DR}
R_i = \sum_{j=1}^{N} h_j(T)v_j - \sum_{j=1}^{N} h_j(1)v_j = \sum_{j\neq i }^{}h_j(T) v_j \quad .
\end{equation}
The last equality holds because we assume that only bank $i$ initially defaults, leading to $h_i(1)=h_i(T)=1$. We define the systemic risk of the entire market as the sum of the individual bank DebtRanks, i.e.
\begin{equation} \label{SR_market}
 R = \sum_{i =1}^{N} R_i \quad.
\end{equation}
\end{defi}
For a motivation of this definition, see also \cite{poledna2016elimination}. For comparison purposes, we also employ a variation of this definition of DebtRank that was presented in \cite{bardoscia2015debtrank}. We refer to this definition as DebtRank2. For more details, see \ref{DR2}. DebtRank2 has been suggested as a micro foundation for shock propagation in networks and is derived from bank balance sheet identities directly. \cite{bardoscia2015debtrank} acknowledge that the original DebtRank formulation can lead to underestimations of systemic risk, because shocks only propagate through a node for a single time and subsequently the node becomes inactive. If a bank receives shocks from different neighbors at sequential times it only transmits the first shock since it becomes inactive after receiving the first shock. Similarly, when a bank receiving a shock is part of a loop and will again receive a shock from the same loop at a later time, it will not forward the shock a second time. The two DebtRanks are the same for tree networks and some other special structures. In general, DebtRank is a lower bound to DebtRank2 \cite[]{bardoscia2015debtrank}. However, since DebtRank2 allows for multiple shock transmissions of a node this leads (in principle) to an infinite number of shocks on networks that contain loops. In practice, the algorithm stops when the shocks become smaller than a predefined value $\epsilon$. However, in the original DebtRank formulation \cite{battiston2012debtrank} point out that an infinite cycling of shocks when loops are present might not be desirable. For this reason and because in the literature the original DebtRank is more widely used, we stick to the original DebtRank for the rest of the paper.  Another interesting generalization of DebtRank is studied by \cite{bardoscia2016distress}, which relaxes the assumption that shocks are propagating linearly.

\section{Minimizing systemic risk as an optimization problem} \label{method}

This section proposes an optimization procedure that rewires a given interbank network to obtain a second (optimal) network that is close to the optimal DebtRank for the prevailing economic environment, i.e. for a given level of equity, bank lending and borrowing, and bank risk.  Because of its recursive definition in Eq. (\ref{DR}), DebtRank is not representable in closed form. This makes it unpractical to use as the actual objective function. Even though an optimization with respect to DebtRank is of course possible in principle, it would be computationally costly, or even infeasible for large networks. We now propose a practical and easy to implement method that is capable of  reducing systemic risk (DebtRank) substantially in empirical networks. For this, we approximate DebtRank, $R$, by a sum of piecewise linear concave functions that then serves as the objective function in the optimization. 

\begin{defi}[Direct Impact]
The direct impact $I_i$ of bank $i$ on its neighbouring banks is defined by 
\begin{equation}
I_i = \sum_{j=1}^{N} W_{ij}v_j =  \frac{1}{\bar{L}}\sum_{j=1}^{N} \min \left( \frac{L_{ij}}{e_j},1 \right){a_j} \quad.
\label{DI}
\end{equation}
The sum of all direct impacts is $I = \sum_{i=1}^{N} I_i$, which can be interpreted as a first-order approximation of the DebtRank.
\end{defi}

Direct impact, $I$, is representable in closed form, Eq. (\ref{DI}), and its special structure allows us to solve the optimization problem with Mixed Integer Linear Programming (MILP) techniques. The optimization procedure rewires links in the network. However, for economic plausibility, it should keep the total assets and liabilities of banks unchanged, as well as the total network volume, $\bar{L}$. These requirements are ensured by corresponding constraints, which have an economic meaning that we discuss in section \ref{constraints}. The optimization problem is now formulated as
\begin{align} \label{eq:optim_simple}
  \min_{L \in \{ M : \; M \in \mathbb{R}_+^{N \times N}, \; M_{ii}=0 \}  } &  \sum_{i=1}^{N}\sum_{j=1}^{N} \min 
  \left( \frac{L_{ij}}{e_j},1 \right)  a_j \nonumber\\ 
    \text{subject to} \qquad & l_i = \sum_{j=1}^{N} L_{ij} \quad , \quad \forall i  \nonumber\\
                   &  a_i = \sum_{j=1}^{N} L_{ji} \quad , \quad \forall i \quad .
\end{align}
The values for $e, l, a, v$, and $\bar{L}$ can be obtained from balance sheets and the interbank network, $L$\footnote{Note that for this optimization only the row and column sums of $L$ are needed, i.e. it can also be performed without the -- usually not accessible -- exact network.}. The objective function is not linear but piecewise linear and concave because of the minimum operator; the sum of concave functions is concave. In Eq. (\ref{eq:optim_simple}) we omit $\bar{L}$ because it is just a positive multiplicative constant. The result of the optimization is the optimal asset-liability matrix, $L^*$. A global optimum exists because of the concavity of the objective function and due to the bounded solution space ($L_{ij} \in [0, \min(a_i, l_i, a_j, l_j)] \; \forall ij $). However, the optimum is not necessarily unique. We find globally optimal solutions by solving an equivalent Mixed Integer Linear Program (MILP), which is derived in the following.

The optimization problem comprises $N^2 - N$ free variables (no self-links), which turns even moderately large interbank markets into large-scale optimization problems. To solve this problem, we linearize the objective function by reformulating it as a Mixed Integer Linear Program (MILP). Since the minimum function is piecewise linear, one can apply standard techniques of mathematical programming to rewrite Eq. (\ref{eq:optim_simple}) as a MILP. 
We use the concept of special ordered sets (SOS), and more specifically, SOS2 constraints for the linearization of the objective function. This concept dates back to \cite{beale1970special} and allows us to find a global solution.

We first provide some intuition of the behavior of the objective function Eq. (\ref{eq:optim_simple}), and then explain the reformulation in detail. Since all $a_j$ are non-negative, we can write them inside the minimum function, and a single term in the objective function in Eq. (\ref{eq:optim_simple}) reads, $\min \big( \frac{ a_j}{e_j} L_{ij}, a_j \big)$. Figure \ref{objectivefunction} in the appendix shows its behavior. It increases until $L_{ij}=e_j$ by $a_j/ e_j$ and remains constant afterwards. Below we show how to relate each entry $L_{ij}$ to a pair of variables, $(y_{2k-1},y_{2k})$. $y_{2k-1}$ accounts for the part of $L_{ij}$, where the objective still increases in $L_{ij}$; $y_{2k}$ accounts for the region, where the objective function is constant (w.r.t. $L_{ij}$). The economic interpretation of the transition point at $L_{ij}=e_j$ is that the liability of bank $i$, with respect to bank $j$ is of the same size as bank $j$'s equity. In the case of $i$'s default, 100\% of $j$'s equity would be destroyed. However, when $L_{ij}>e_j$,  $y_{2k} = \min(0, L_{ij}-e_j)$ does not affect the objective function anymore, since more than 100\% of $j$'s equity cannot be consumed. Note that the remaining loss of $\min(0, L_{ij}-e_j)$ is born by creditors of $j$, which are outside the interbank system. We now show more formally how the objective function can be transformed into a MILP with the help of the variables $y$ and a set of dummy variables, $\delta$.

We stated the optimization problem in matrix terms. In numerical optimization it is more common to optimize over vectors.  We therefore rewrite $L \in \mathbb{R}_+^{N \times N}$ into a vector $x\in \mathbb{R}_+^{N^2}$ by stacking the columns of $L$, 
\begin{equation}
x = \text{vec}(L)= (L_{11},\dots,L_{N1}, L_{12}\dots,  L_{N2}, L_{1N}, \dots,L_{NN})^\top \quad.
\end{equation}
Note that for ease of notation and implementation we keep the diagonal entries $L_{ii}$, $\forall i$. Similarly, we define vectors of length $N^2$ for representing assets, liabilities, and equities,
\begin{equation}
\bar{e} = (\underbrace{e_1, \dots, e_1}_{N \, {\rm times}},e_2, \dots,\underbrace{e_N, \dots, e_N}_{N \, {\rm times}})^\top \quad ,
\end{equation}  
\begin{equation}
\bar{a} = (\underbrace{a_1, \dots, a_1}_{N \, {\rm times}},a_2, \dots,\underbrace{a_N, \dots, a_N}_{N \, {\rm times}})^\top \quad, 
\end{equation}  
and 
\begin{equation}
\bar{l} = (\underbrace{l,\dots, l}_{ N \, {\rm times}})^\top  \quad.
\end{equation}
Now we can write the objective function in  Eq. (\ref{eq:optim_simple}) as
\begin{eqnarray} \label{eq:optim_vector}
\min_{x \in \mathbb{R}_+^{N^2} } &   \sum_{j=1}^{N^2} \min \left( \frac{ \bar{a}_j}{\bar{e}_j} x_{j},\bar{a}_j \right) \quad .
\end{eqnarray}
The elements in $x$ corresponding to the diagonal elements of $L$ have to be zero, which can be enforced with additional constraints, or directly in the optimization software. To translate the objective function into a linear form $c^\top y$, every variable $x_i$ is split into two parts, 
 $y_{2i-1} $ and $y_{2i}$, with $x_i = y_{2i-1}+y_{2i}$, and
\begin{eqnarray}
y_{2i-1} & = & \min(x_i, \bar{e}_i) \quad, \\
y_{2i}   & = & \min(x_i - \bar{e}_i, 0) \quad .
\end{eqnarray}
The first part, $y_{2i-1}$, indicates the range of $x_i$, where an increase $\Delta x_i$ leads to an increase of the objective function by $\Delta x_i (\bar{a}_i/\bar{e}_i)$. At $x_i=\bar{e}_i$, the objective function does no longer increase with $x_i$. This range of $x_i$ is accounted for by $y_{2i}$. To reformulate the objective function in terms of the new variables $y$, we need to introduce a  vector of binary variables $\delta \in \{0,1\}^{2n^2}$ in the following way 
\[ \delta_j = \begin{cases}
1 \qquad \text{if} \quad y_j > 0 \\
0 \qquad \text{if} \quad y_j = 0 \quad .
\end{cases} \]
With $\delta$ we can formulate the following constraints for the pairs $(y_{2i-1},y_{2i})$, for all $i$, 
\begin{eqnarray} 
 \delta_{2i-1} & \geq & \delta_{2i} \label{eq:con_delta1} \\
 y_{2i-1} & \geq & \delta_{2i} \bar{e}_i \label{eq:con_delta2} \\
 y_{2i-1} & \leq & \delta_{2i-1} \bar{e}_i \label{eq:con_delta3} \\
 y_{2i}   & \leq & \delta_{2i} \max \left(0, \min \left( \bar{a}_i,\bar{l}_i \right)- \bar{e}_i \right) \quad . \label{eq:con_delta4}
\end{eqnarray}
Constraints (\ref{eq:con_delta1}) - (\ref{eq:con_delta4})  ensure the equivalence of the reformulated problem in Eq. (\ref{eq:optim_milp}), and the original problem in Eq. (\ref{eq:optim_simple}). In particular, Eq. (\ref{eq:con_delta1}) enforces that $y_{2i}$ can only be larger than zero if $y_{2i-1}$ is larger than zero. Constraints (\ref{eq:con_delta2}) and (\ref{eq:con_delta3}) enforce that $y_{2i-1}$ must be smaller than $\bar{e}_i$, and that if $y_{2i}$ is bigger than zero, $y_{2i-1}$ has to be equal to $\bar{e}_i$. Finally, Eq. (\ref{eq:con_delta4}) ensures that $x_i = y_{2i-1}+y_{2i}$ is smaller than the respective row and column sum of the corresponding entry in the liability matrix. We finally define the vector $c$ of length $4N^2$, which determines the slope with which the respective entries in $y$ increase,   
\begin{equation} \label{eq:c}
  c_j = \begin{cases}
  \frac{\bar{a}_i}{\bar{e}_i} \qquad \text{if} \; j=2i-1 \; \text{, and} \; i \leq N^2 \\
  0 \qquad \; \; \text{if} \; j= 2i \; \text{, and} \; i \leq N^2  \\
  0 \qquad \; \; \text{if} \; 2N^2 < j \leq 4N^2 \quad .
  \end{cases}
\end{equation}
It follows that every second entry in $c_{2i}$ is equal to zero, since the even components $y_{2i}$ are not increasing the objective function. They correspond to the part of $x_i$, where the objective function is capped to $\bar{a}_i$. The odd parts, $c_{2i-1}$, represent the slopes. The last $2N^2$ zeros ensure that the binary variables $\delta$ do not affect the value of the objective function. The objective function can now be written as $c^\top z$, where $z = (y,\delta) \in \mathbb{R}^{4n^2}$.

The constraints for $\delta$ and $y$, Eqs. (\ref{eq:con_delta1}) -- (\ref{eq:con_delta4}), are compactly reformulated as $A_1z = 0$, where $A_1 \in \mathbb{R}^{4n^2 \times 4N^2}$, and $0$ denotes a zero-vector of length $4N^2$. The constraints on the row and column sums of the liability matrix in the initial problem, Eq. (\ref{eq:optim_simple}), can be written in standard matrix form as $A_2 z = a$, and $A_3z = l$. $A_2$ and $A_3$ are  $N \times 4N^2$ dimensional matrices consisting of zeros and ones\footnote{Note that there is at least one redundant equation in this set of linear constraints, since $N$ column sums and $N-1$ row sums imply the $N$th row sum.}. The exact structure of the constraint matrices $A_1,A_2$, and $A_3$ is outlined in \ref{app:example}. Finally, the optimization problem of Eq. (\ref{eq:optim_simple}) as a MILP reads  
\begin{eqnarray} \label{eq:optim_milp}
 \min_{z \in \mathbb{R}_+^{4N^2}} &  c^\top z & \\
 \text{subject to}  \	& A_1z & \leq 0 \\
                  		 &  A_2z & = a \\
                  		  & A_3z & = l \quad . 
\end{eqnarray}
This method is generic and is generally applicable to all direct financial exposure networks, where systemic risk is quantified by DebtRank. If different types of financial network are considered, the liability matrix $L$ has to be replaced with the corresponding exposure matrices. Depending on the financial network type, various further constraints can be considered to ensure that certain economic properties of individual banks (which depend on the network but should kept constant) indeed remain the same after the optimization. We continue by discussing such constraints in more detail.
 
\subsection{Implementing economic constraints} \label{constraints}

As mentioned, the constraints in Eq. (\ref{eq:optim_simple}) ensure not only that banks retain their size after optimization, but they also have an important economic interpretation. Row and column sums represent the interbank liabilities and interbank assets of each bank. Keeping these constant implies that each bank retains its amount of liquidity\footnote{Since we deal only with a single liability matrix, $L$, we implicitly assume in the optimization procedure that all liabilities have the same maturity, which is of course not realistic. If a family of matrices, $L_1, \dots, L_t$, describing the interbank liabilities for various maturities (or maturity buckets) $1,\dots, t$ is available, and the optimization procedure is applied to each maturity separately, then the original maturity structure is unaffected.} from the interbank market after optimization. If we assume that the liquidity a bank requires from the interbank market originates from its operational business, it is important that this activity is not distorted by the optimization procedure. 

Another important type of economic constraint is related to economic risk. In direct exposure networks counter-party credit risk plays an important role when making lending decisions. In the case of indirect exposure networks, such as overlapping portfolio risk, risk associated to the financial assets, which are held by the financial institutions play a crucial role when making investment decisions. The most important types are credit, market, and interest rate risk. For the sake of comparability of empirically observed reference networks and optimized networks, it is desirable to have constraints, which ensure that the risks for the individual institutions remain comparable before and after optimization. 

For interbank networks none of the lending banks should end up with a higher counter-party credit risk after the optimization. To achieve this, we introduce another linear constraint to ensure that the credit risk in all interbank loan portfolios is approximately maintained. This constraint accounts for individual economic conditions of banks, which are affected by the network structure. We aim to model this feature by fixing the predominant credit risk weighted exposure of the interbank loan portfolio of each bank. 

Let the considered credit risk indicator of bank $i$ be $\kappa_i$. For a given liability matrix $L$, the risk weighted interbank loan exposure of bank $j$, that is implied by the interbank network $L$, is then given by $ r_j = \sum_{i=1}^{ N } L_{ij}\kappa_i$, or in matrix notation, $r = L^\top \kappa$ . To include this constraint in the MILP of Eq. (\ref{eq:optim_milp}) we need to translate $r$ to 
\begin{equation}
A_4 z = L^\top \kappa \quad . 
\label{eq:constraint4}
\end{equation}
Details of the matrix $A_4$ are found in \ref{app:example}. We explain in \ref{equivalence}  that the formulation of constraint Eq. (\ref{eq:constraint4}) as equality and smaller or equal yield the same optimal value of the problem in Eq. (\ref{eq:optim_milp}), given that the row and column sum constraints are in place. 
Further, this constraint also keeps the earnings from the interbank loan portfolio similar before and after optimization, because the interest rate earned on an interbank loan should strongly reflect the credit risk level of the borrower. Additionally, also the regulatory capital levied on the interbank loan portfolio remains comparable, because capital requirements depend on the risk weighted assets of the respective bank. Since the risk weighted interbank loan exposure remains constant in the optimization, also the risk weighted assets should retain their size. 

For the case of optimizing indirect exposure networks, similar risk constraints can be implemented. For example, \cite{pichler2018systemic} consider Markowitz mean-variance conditions for optimizing financial exposures emerging from common asset holdings and discuss further possible constraints. Other meaningful constraints for financial asset networks are credit risk constraints, such that the average credit risk -- of e.g. a bond portfolio -- remains comparable. To keep the interest rate risk of fixed income portfolios similar across the optimization, another linear constraint can account for the maturity or duration of the assets. In general, different financial networks will require different economic constraints.

\section{Optimization of empirical Austrian interbank networks}\label{data}

The solution to the MILP yields a network with minimal direct impacts, $I$, but not necessarily one with minimal systemic risk in terms of DebtRank, $R$. However, our computations demonstrate the great effectiveness of this approximation in massively reducing overall systemic risk.

 \begin{figure}[t]
 	\vspace{-2cm}
 	\centering
 	\includegraphics[scale=.55, keepaspectratio]{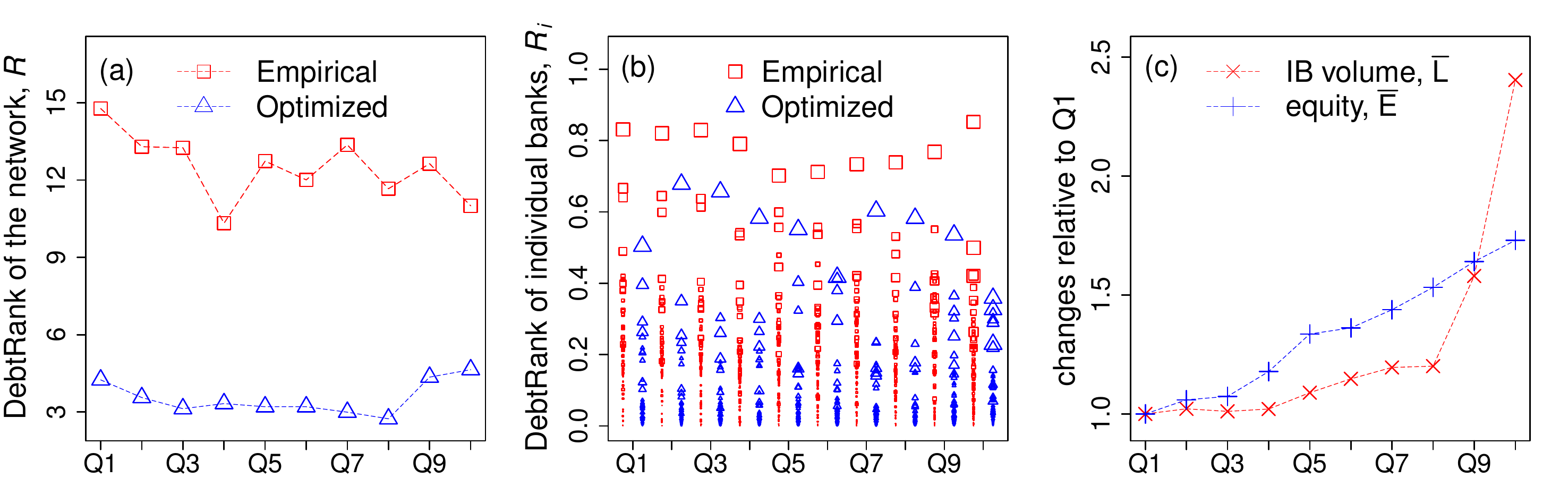}
 	\caption{(a) Total DebtRank, $R$, of the empirical Austrian interbank networks across 10 quarters from 2006 to 2008 (squares). 
	For the optimized networks the DebtRank is drastically reduced (triangles). 
	The optimization reduces systemic risk (measured in DebtRank) by a factor of approximately 3.5. 
	(b)  Individual DebtRank, $R_i$, of 70 banks for the empirical and optimized networks in the respective quarters. 
	Here symbol sizes are proportional to the banks' interbank liabilities, $l_i$. 
	We see that typically large banks have high $R_i$, however note that there are many exceptions with small banks having considerable 
	systemic risk. 
	(c) Total interbank market volume, $\bar{L}$, and equity, $\bar{E}= \sum_{i=1}^{ N }e_i$, over the ten quarters.
	While decreasing in the first eight quarters, the ratio $\bar{L}/\bar{E}$ increases substantially in Q9 and Q10.
	   }
 	\label{drempvsmin}       
 \end{figure}

We apply the optimization to a data set consisting of 10 snapshots of Austrian interbank networks at 10 quarters from 2006 to 2008. The Austrian interbank network has been studied before by e.g. \cite{boss2004network}, \cite{elsinger2006risk}, \cite{caccioli2015overlapping}\footnote{\cite{caccioli2015overlapping} uses the same data set that originally consisted of 12 quarters.  Due to obvious data errors we dropped 2 of the 12 observations.}. The sample contains between 824 and 846 banks. The  Austrian banking system accommodates many very small cooperative banks, which can not be considered as systemically important. We use the 70 largest banks with respect to total assets in the corresponding quarter for numerical feasibility. These account for about 86\% of total assets; the 70th largest bank accounts for around 0.12\% of total assets. The 70 banks with the largest total assets cover around 71\% of the interbank market. We choose the banks' total asset size as the selection criterion because total assets should be a more stable quantity than interbank market shares. We deal with a fully anonymized data set, which makes it impossible to estimate the bank's credit risk indicators, $\kappa_i$. Approaches for estimating  $\kappa$ are outlined in Section \ref{creditrisky} for the case that sufficient data would be available. For demonstration purposes we approximate $\kappa$ by the leverage ratio of the banks 
\begin{equation}
\kappa_i = \frac{\text{\rm total \, assets}_i}{\text{total \, assets}_i - \text{total \, liabilities}_i} \quad .
\end{equation} 
We assume that a higher leverage ratio implies higher credit risk.
 
To solve the optimization problem numerically, we employ the MILP solver cplex, available in the R Optimisation Infrastructure (ROI) package \cite[]{theussl2017roi}. The optimization can be performed on a standard notebook and takes between a few minutes to several hours, depending on the network sample.  

\begin{figure}[t]
	\vspace{-2cm}
	\centering
	\includegraphics[scale=.55, keepaspectratio]{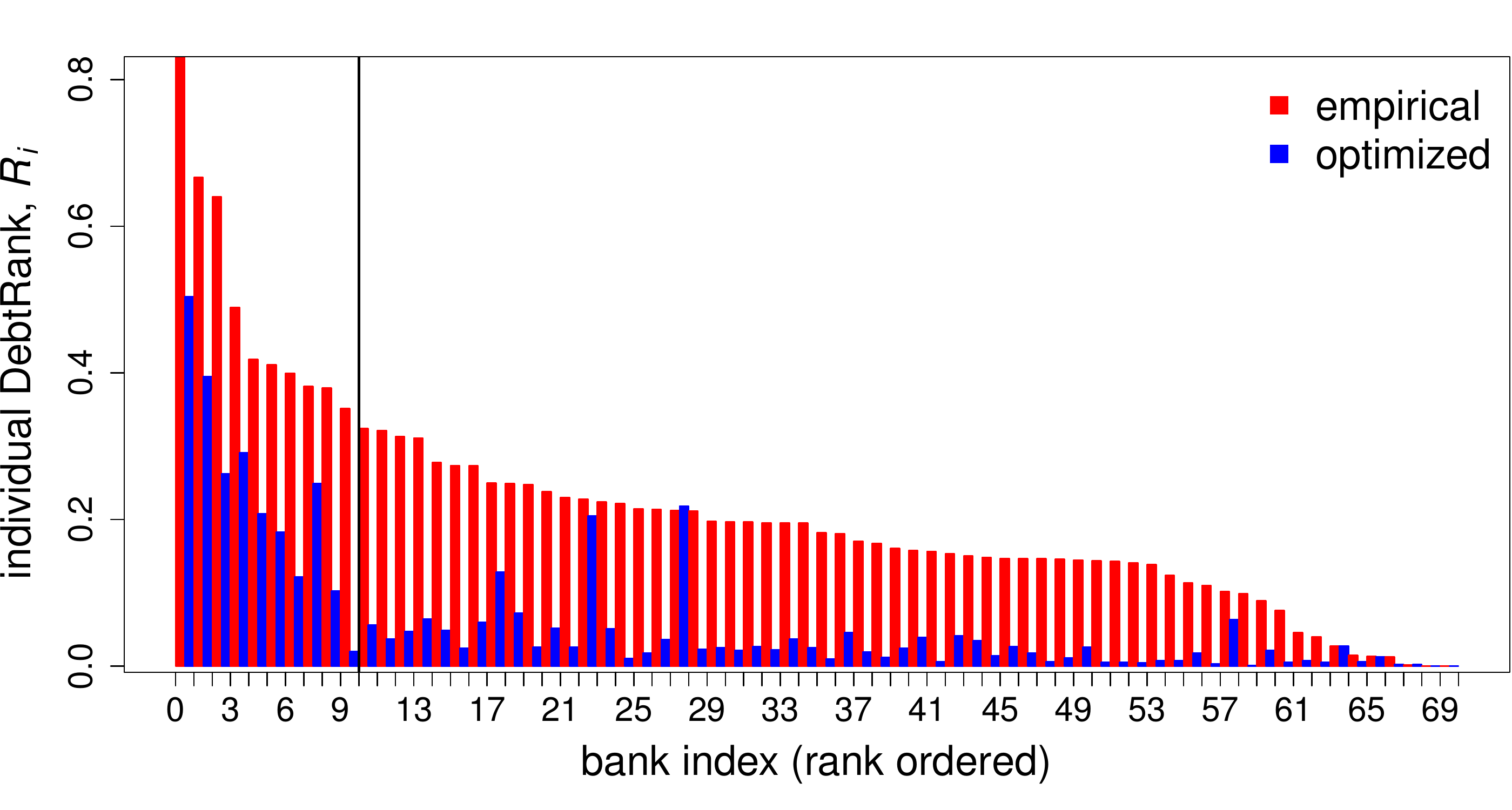}
	\caption{Systemic risk profile (DebtRank $R_i$) of the 70 banks for the empirical (red) and the minimized (blue) 
	networks in quarter Q1. Banks are rank ordered with respect to their DebtRank, $R_i$, in the empirical network. 
	It is visible that systemic risk is drastically reduced for practically all banks, with only one exception. 
	For the 10 most risky banks a reduction of DebtRank, $R_i$ by a factor of 2.1  is observed, for higher ranks, 
	the reduction by a factor of 5.1 is even more drastic. Similar results hold for the other quarters.
	}
	\label{drcomparison}       
\end{figure}

\subsection{Results}\label{results}

The reduction of systemic risk obtained by the optimization procedure is summarized in Figure \ref{drempvsmin} (a). The values of the total DebtRank, $R$, after optimization are substantially lower than the corresponding empirical ones across all quarters. The average DebtRank in the empirical and optimized networks are around 12.51 and 3.54, respectively, meaning that the average total DebtRank reduction amounts to approximately 71\%, or a factor of roughly 3.5. Figure \ref{drempvsmin} (b) shows the individual DebtRanks, $R_i$, of the 70 banks for the empirical case (red squares) and the optimized (blue triangles). The size of the symbols represents the banks' interbank liabilities, $l_i$. The figure shows two facts. The first is that in the optimized network at least one bank always remains relatively systemically risky with respect to the bulk of banks, even though their DebtRank is substantially reduced. The second observation is that the DebtRank reduction for small and medium sized banks, indicated by triangle size, seems to work even better than for the large banks. Figure \ref{drsizecomparison} shows the relationship of DebtRank $R_i$ and interbank liabilities $l_i$ in more detail. In the empirical networks small banks severely ``punch above their weight'', i.e.  banks with small interbank liabilities frequently have high DebtRanks, $R_i$, and their default would cause -- judging by their size -- unnecessary systemic events. The optimization remedies this problem and renders banks with small interbank liabilities systemically negligible.  

In Figure \ref{drempvsmin} (a) it is seen that from Q8 to Q10 the optimized DebtRank increases, while the empirical DebtRank continues its downward trend. To understand why, in Figure \ref{drempvsmin} (c) we show the total interbank market volume and the total equity in the system over time, relative to the values in Q1. Larger levels of  equity---all other things kept equal---should reduce DebtRank, and an increase of the market volume should increase DebtRank. Thus, the sharp increase of the market volume, $\bar{L}$, from Q8 to Q10 could be the explanation for the observed increase in the optimized DebtRank. 

\begin{figure}[t]
	\centering
	\includegraphics[scale=.5, keepaspectratio]{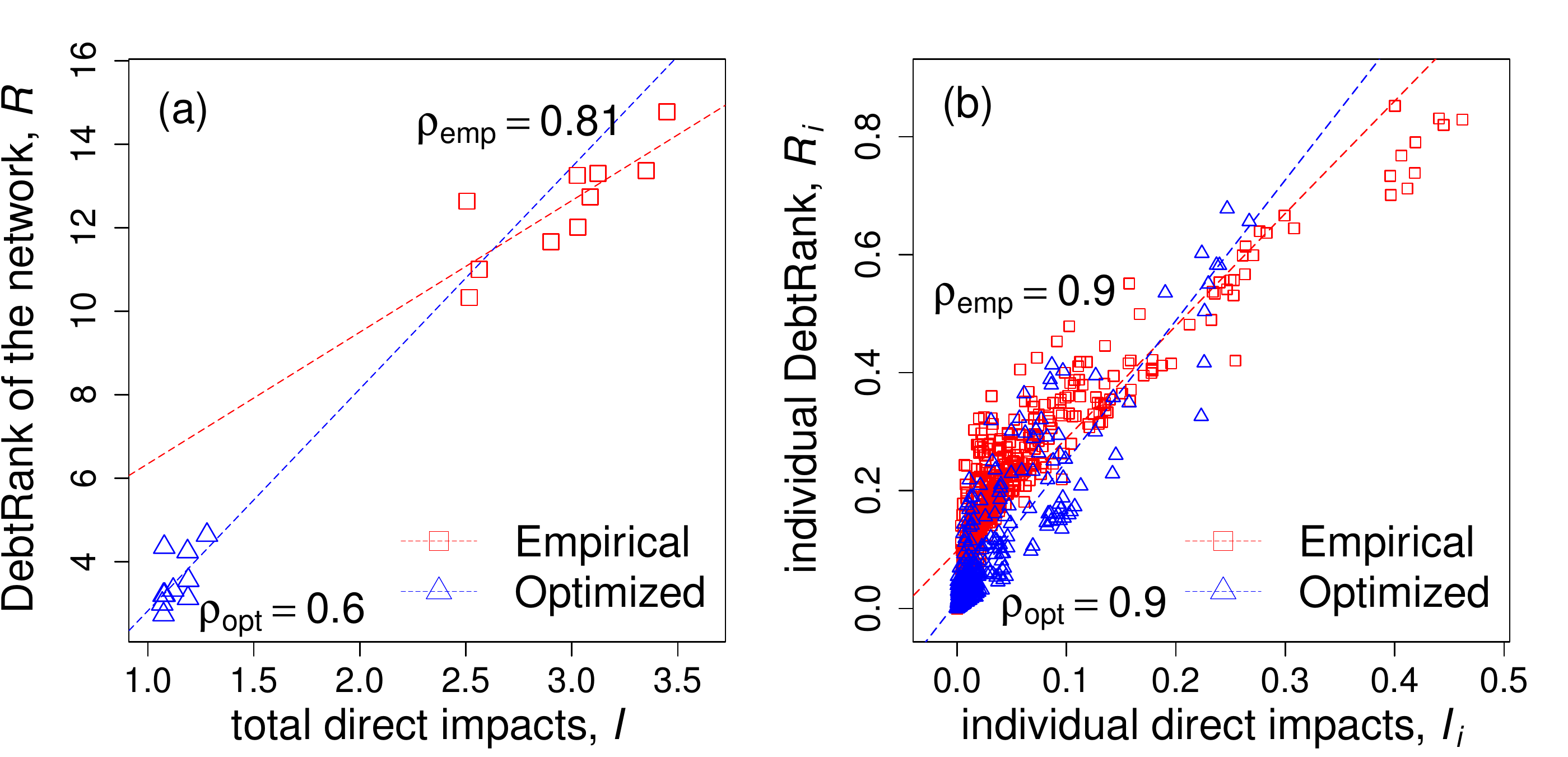}
	\caption{(a) Total DebtRank values, $R$, versus direct impacts, $I$, for the 10 quarters of the 
	empirical (squares) and the optimized networks (triangles). 
	The correlation coefficients of $R$ and $I$ for the empirical and optimized 
	cases are $\rho_{\rm emp}=0.81$, with a p-value of $0.004$, and $\rho_{\rm opt}=0.6$, with a p-value of $0.07$, respectively.
	 (b) Shows the same comparison on the individual bank level, $R_i$ versus $I_i$, with $\rho_{\rm emp}=0.90$, and $\rho_{\rm opt}=0.9$. 
	 The associated p-values are smaller than 2.2e-16. The dashed lines are obtained by simple linear regression. 
	 }
	\label{divsdr}	
\end{figure}

Figure \ref{drcomparison} depicts the systemic risk profile for Q1, where $R_i$ is shown for 70 banks of the empirical and the optimized case. Banks are ordered according to their empirical DebtRank, $R_i$; the most systemically risky institution is shown to the very left. The effectiveness of the optimization is clearly seen. DebtRank levels are decreased substantially for almost all 70 banks, with two exceptions, where banks have a slightly higher DebtRank after the optimization. For the 10 most risky banks (left of vertical line) DebtRank is reduced by a factor of around 2.1, for higher ranks, the reduction is even more pronounced, and amounts to a factor of 5.1. For most banks, DebtRank is decreased to marginal levels. Similar observations hold true for all quarters; in some a DebtRank reduction is achieved for all 70 banks. 

Figure \ref{divsdr} shows the relation of our objective function (direct impacts), $I$, and DebtRank, $R$, that serves as our measure of systemic risk, with which we also judge the effectiveness of the optimization. On the network level, the total DebtRank and direct impacts of the empirical networks are linearly related with a correlation coefficient of $\rho_{\rm emp}=0.81$, and a $p$-value of $p_{\rm emp}=0.004$. This confirms {\em a posteriori}  that minimizing the direct impacts is indeed a reasonable and effective way to minimize DebtRank. In the optimized networks the linear relationship is weaker ($\rho_{\rm opt}=0.6$, and a $p$-value of $p_{\rm opt}=0.07$). This indicates that the optimization achieves a stronger reduction in direct impacts than in DebtRank. Figure \ref{divsdr} (b) shows the same situation for the bank's individual levels of DebtRank, $R_i$, and direct impacts, $I_i$. The linear correlations for both network types are higher ($\rho_{\rm emp}=\rho_{\rm opt} = 0.9$) and their $p$-values are below 2.2e-16. 
The respective results for DebtRank2 (\cite{bardoscia2015debtrank}) are shown in \ref{DR2}. Here, the optimization achieves an average reduction of DebtRank2 of about 15\%.

\subsection{How networks change during optimization}

\begin{figure}[H]		
	\centering
	\includegraphics[scale=.38, keepaspectratio]{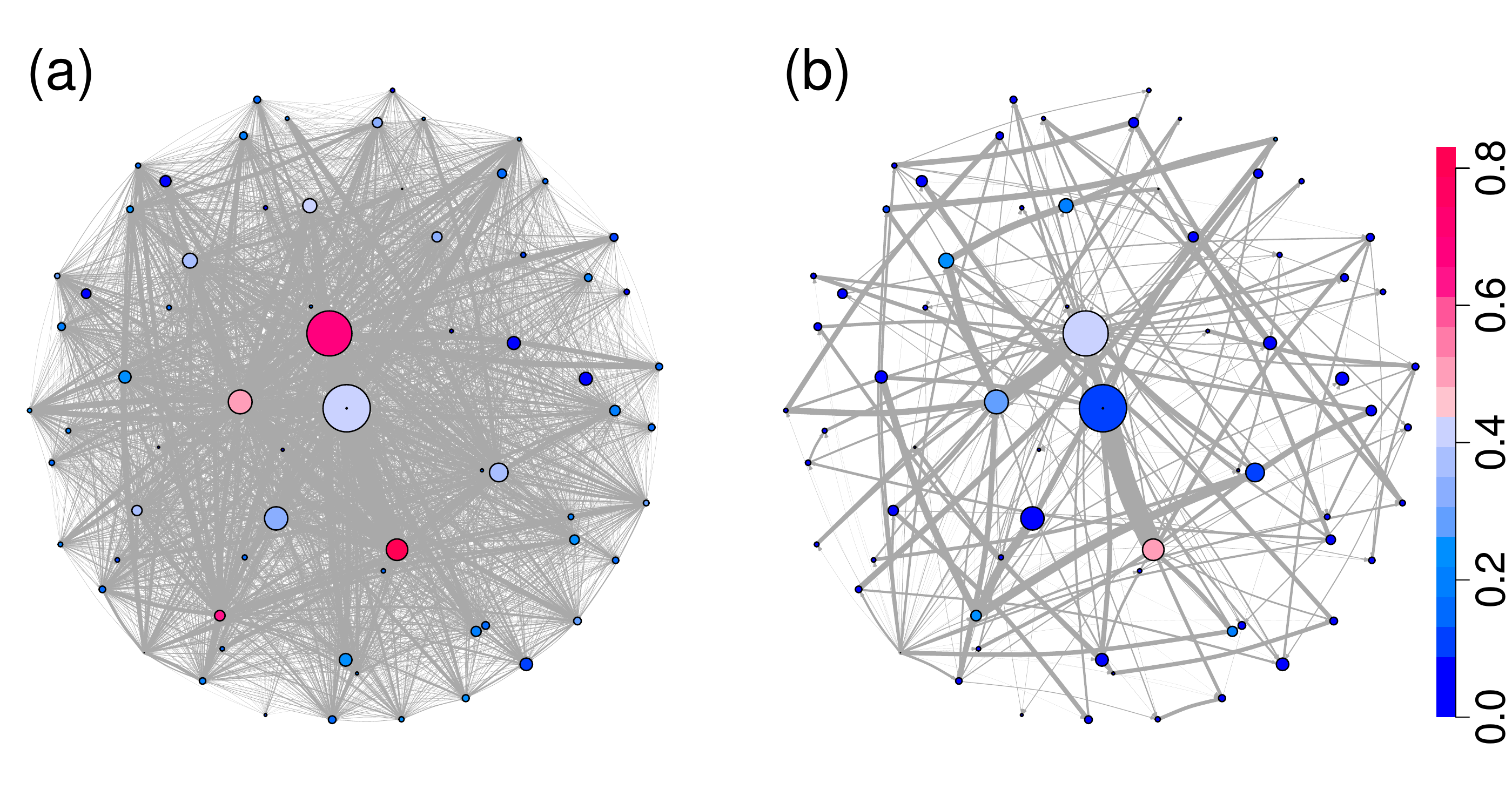}		
	\includegraphics[scale=.38, keepaspectratio]{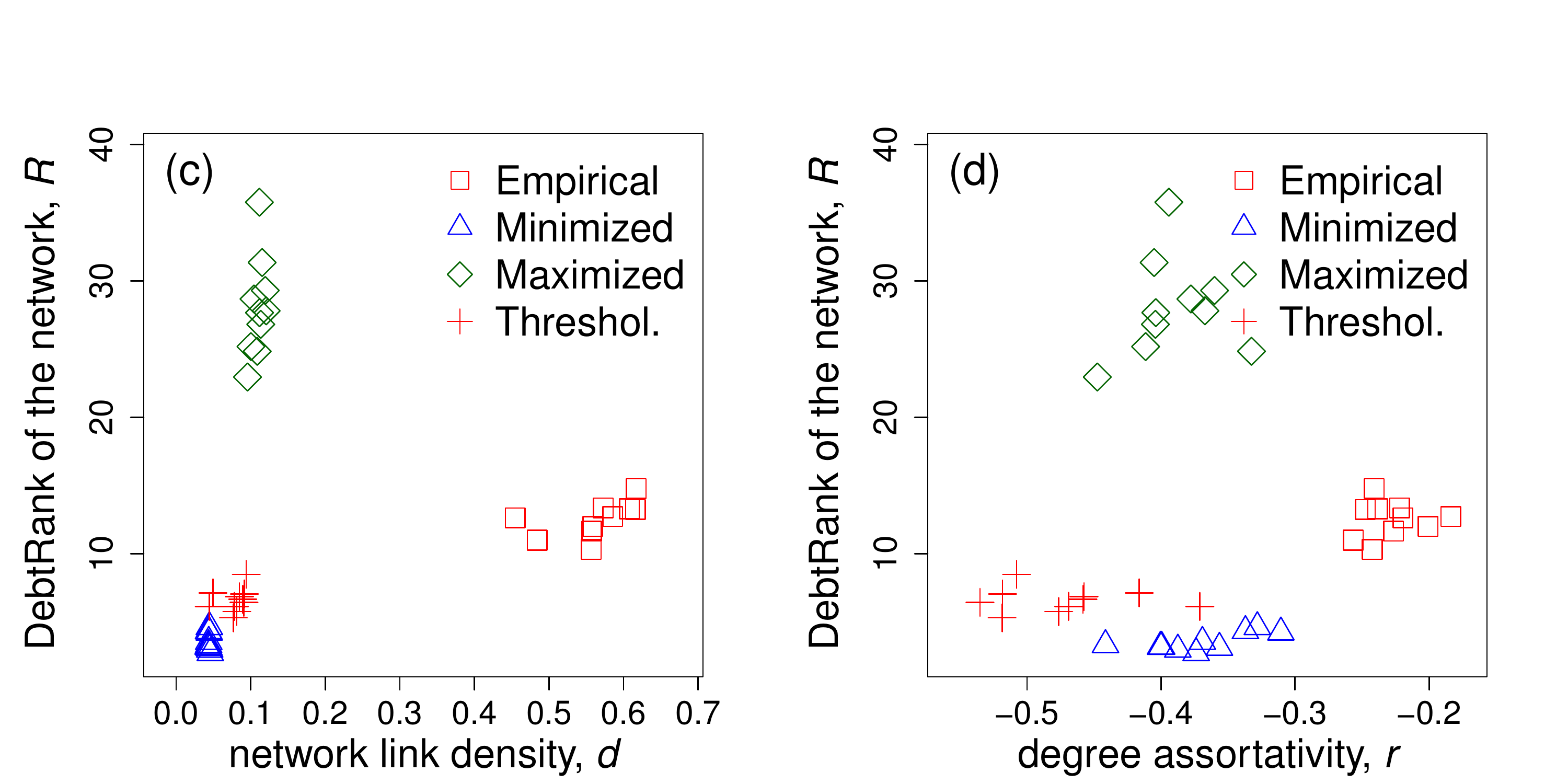}
	\includegraphics[scale=.38, keepaspectratio]{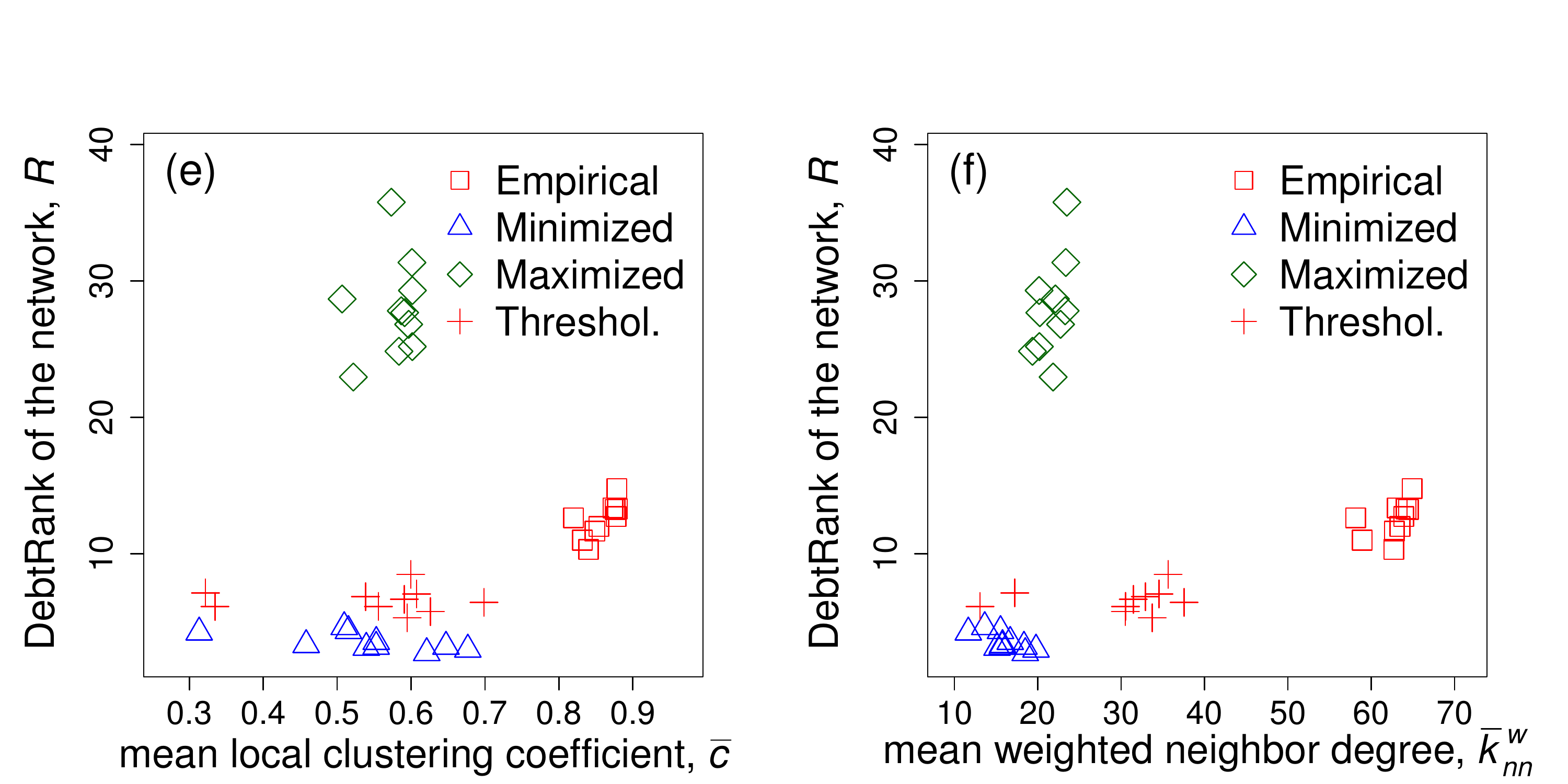}		
	\caption{Interbank networks  before and after optimization. 
		(a) Empirical asset-liability network, $L$, as in Q1, in comparison to (b) the optimized network, $L^*$. 
		Node colors of banks represent their DebtRank (large DebtRank is red, small is blue). Node size is proportional to equity, $e_i$. 
		It is obvious that the optimized network is considerably sparser. 
		(c) DebtRank of the empirical, the minimized, and the maximized networks plotted against the networks' link 
		densities, $d$. Every symbol represents a quarter. It is visible that sparse networks can have both, high and low DebtRank. 
		(d) DebtRank, $R$, against the degree-weighted assortativity, $r$. 
		We see a similar level of dis-assortativity in the maximized and minimized networks, 
		while the empirical network is more assortative, and the thresholded empirical network is less assortative.
		(e) DebtRank, $R$, plotted against the mean local clustering coefficient, $\bar{c}$. 
		We see that there is a tendency towards higher local clustering in the empirical network. 
		The minimized and the thresholded networks show similar average clustering. 
		The average clustering in the maximized networks is slightly higher than in the minimized networks. 
		(f) DebtRank, $R$, plotted against the average weighted nearest neighbor degrees,  $\bar{k}_{nn}^w$. 
		The smallest values are observed for the minimized network, followed by the maximized networks. 
		The thresholded networks exhibit higher, the empirical networks substantially higher values.  
	}
	\label{densityassortativity}  	
\end{figure}

Figure \ref{densityassortativity} (a) shows the original interbank asset-liability network $L$ before the optimization for quarter Q1. The case after optimization is seen in (b). The nodes represent banks, size is the banks' equity, the colors represent the DebtRank value (dark red is high, light tones are medium, dark blue is low $R_i$). There are obvious differences. We now ask how the topology of interbank networks changes due to the optimization process. The average degree of the minimized network (from the binary adjacency matrix) is $ \bar k=3.04$
versus the empirical network $ \bar k=38.71$.

The in- and out-degree distributions for the different network types pooled together for all ten quarters are provided in Figure \ref{inoutdeg_dist} (a) and (b). The average in- and out- strength of the networks are unchanged, due to the constraints that keep $a_i$ (in-strength) and $l_i$ (out-strength) fixed. 

The most prominent observation is that networks after the optimization become sparser. Figure \ref{densityassortativity} (c) shows that the minimized 
network (blue triangles) is extremely sparse with an average link density of around 4.4\%. Every dot represents one of the 10 quarters. The link density of the network (connectancy) is defined as the fraction of links being present in the network, $d = {m}/{(N(N-1))}$, where $m$ is the number of present links and $N(N-1)$ is the number of possible links. In contrast, the empirical networks (red squares) exhibit an average link density of approximately 56\%. Note however, that by slightly thresholding the empirical networks, link densities of about 10\% are obtained, see \ref{topology}. One could be led to believe that high link density is related to high DebtRank. This is not necessarily true. To show this, we computed the maximum direct impact networks (where we maximize Eq. (\ref{eq:optim_milp})), which leads to networks with higher DebtRank than the empirically observed ones. Interestingly, these maximized networks (green diamonds) are also sparse, with an average link density of 11\%. The maximized and thresholded networks are visualized in Figure \ref{empnwplot}. Sparse networks can have low or high DebtRank. Also \cite{krause2019} find for their optimization that the minimized and maximized networks are considerably sparser than the original network. For a sample of 53 banks they report an average degree of $\bar{k}= 14.4$ for the reference network and an average degree of $\bar{k}= 2$ for their minimized and maximized networks each. \cite{aldasoro2017bank} find for a network of 20 banks (obtained from a theoretical model, where banks lend/borrow in an optimal way w.r.t. their utility function) a link density of around 7.3\%.

In Figure \ref{densityassortativity} (d) DebtRank $R$ is plotted against the degree assortativity, which is calculated as
\begin{equation}
r = \frac{\sum_{i}^{}j_i k_i - m^{-1} \sum_{i}\sum_{i'} j_i k_{i'}}{
	\sqrt{ \Big(\sum_{i}j_i^2 - m^{-1} \big( \sum_{i}j_i \big)^2 \Big) \; 
		   \Big(\sum_{i}k_i^2 - m^{-1} \big( \sum_{i}k_i \big)^2 \Big)}} \quad ,
\end{equation}
where $j_i$ is the excess in-degree and $k_i$ the excess out-degree of the nodes, which are at the beginning and the end of link $i$, and $m$ is the number of all links \cite[Eq. (26)]{newman2003mixing}. We report the average of the ten assortativity coefficients for the four network types: empirical $-0.23$, minimized $-0.37$, maximized $-0.39$ and thresholded $-0.47$.  We see that the minimized and maximized networks tend to have a similar degree assortativity, whereas the empirical networks are more assortative, and the thresholded network more disassortative.  

A similar picture is found for the clustering coefficient, $\bar{c}$. We compute it from the unweighted undirected (symmetrized) adjacency matrix for each of the ten quarters. The clustering coefficient of the network is defined as $\bar{c} =  1/N \sum_{i=1}^{N} C_i$, where the local clustering coefficient $C_i$  of node $i$ is defined as the number of connected pairs of neighbors of $i$, divided by the number of pairs of neighbors of $i$. Figure \ref{densityassortativity} (e) plots the clustering coefficients $\bar{c}$ for each quarter against the DebtRank, $R$, of the respective quarter.  The average of the clustering coefficients for the ten quarters is $0.86$ for the empirical, $0.54$ for the minimized, $0.58$ for the maximized, and $0.54$ for the thresholded network. Again, the average clustering coefficient for the minimized and maximized networks show a similar behavior. To give a more detailed picture of the clustering behavior, we show the histogram of local clustering coefficients, $C_i$, for the different network types pooled for all ten observations in Figure \ref{inoutdeg_dist} (c). Approximately in line with these numbers, in their theoretical model \cite{leduc2017incentivizing} report average clustering coefficients for 500 networks of about $\bar c \sim 0.7$, for networks with usual systemic risk levels, while for networks that are obtained under a systemic risk tax (that systematically reduces systemic risk by incentivizing agents), they find $\bar c \sim 0.3$.

Finally, in Figure \ref{densityassortativity} (f) we study the average weighted nearest neighbor degree, $\bar{k}_{nn}^w = \sum_{i=1}^{N}k_{nn,i}^w$, where $k_{nn,i}^w$ is defined as $k_{nn,i}^w = (a_i+l_i)^{-1} \sum_{j=1}^{N}(L_{ij}+L_{ji})k_j$. The degree of the neighbors of node $i$ are weighted with the size of the mutual exposure between them and are standardized by the sum of $i$s' interbank liabilities and assets. We report the values for the empirical $63$, minimized $16$, maximized $21$, and thresholded $30$ networks. For the mean weighted nearest neighbor degree a notable difference between the minimized and maximized networks is observed. However, both types exhibit much smaller values than the empirical networks. The distribution of $k_{nn,i}^w$ across all ten observations is shown in \ref{inoutdeg_dist} (d).

These findings in the network topology measures clearly indicate that there are substantial differences between the optimized and the empirical networks. Interestingly, the topology measures for minimized and maximized networks are often similar.
Not unexpectedly, due to the the large number of small exposures in empirical networks, the thresholded networks often appear more similar (based on various network measures) to the optimized ones than to the empirical ones.

\section{Discussion and Conclusion} \label{conclusion}

The purpose of this paper was to propose a straightforward method for quantifying the systemic risk reduction potential in financial exposure networks. We are able to compute the (approximate) mimimum systemic risk which is theoretically attainable in a financial exposure network under given economic constraints. In summary, the proposed optimization approach leads to a drastic reduction of systemic risk (DebtRank) in interbank networks, while keeping the overall economic conditions of banks (their total assets, liabilities, and average risk) practically unchanged. The obtained optimal financial networks can be used as a best-case benchmark for systemic risk in empirical financial networks with a given total market volume, equity, assets, liabilities, and possibly other constraints. It allows us to estimate the maximum systemic risk reduction potential, and may serve as a benchmark quantity to monitor the divergence of actual markets from- or convergence to their systemic risk optimum. 

In the literature, reducing systemic risk in financial markets is typically discussed in the context of introducing higher capital requirements. Since systemic risk is inextricably linked to the topology and details of the underlying financial networks, the question arises of how much systemic risk can be taken out of the system by reorganizing the underlying financial networks. First contributions in this direction were \cite{thurner2013debtrank}, \cite{poledna2016elimination}, \cite{leduc2017incentivizing}, and \cite{pichler2018systemic}. Here we have shown that the network-based systemic risk reduction potential is potentially huge, when compared to what can be gained from increased capital surcharges, \`{a} la Basel III, \cite[]{poledna2017basel}. We showed in detail how the quantification of minimal systemic risk can be achieved by a reorganization of the financial network with Mixed Integer Linear Programming (MILP). The optimization can be performed with readily available MILP solvers.

We demonstrated the effectiveness of the proposed method by applying it to a data set consisting of 10 quarterly observations of the Austrian interbank market, where we focus on the 70 largest banks in the sample, which cover about 86\% of total assets, and 71\% of the interbank assets. The optimization shows that DebtRank can be reduced on average by a factor of 3.5, under constraints that guarantee that the banks' interbank assets and liabilities, as well as their leverage ratio weighted interbank loan exposure are kept constant. We find evidence that the largest players in the interbank market still remain systemically relevant after optimization (even though much less so), whereas the DebtRank values for most small and medium sized banks are reduced to levels, which are systemically negligible. In the empirical networks small banks often ``punch above their weight'' with respect to systemic risk, i.e. their default is causing disproportionately large systemic losses. The optimization remedies this problem, such that the systemic risk caused by banks is approximately proportional to their size. In other words, the problem of `too central to fail'  can be addressed by reorganizing the underlying network topologies, while the problem of `too large to fail' remains, at least to a certain extent. It is reasonable to conclude that  for systemic risk-efficient allocations (i.e. the optimized networks) the individual systemic risk becomes
more proportional to liabilities.

Our findings highlight the role of financial network topology to understand and substantially reduce systemic risk. Policies that explicitly account for the network structure of financial markets are a necessary and efficient way to reduce systemic risk. An obvious policy proposal has already been made in \cite{poledna2016elimination} and \cite{leduc2017incentivizing}, where it is suggested to tax (or insure) the externalities of systemically risky transactions and thus introduce an incentive scheme to avoid those; in effect, the network topology changes towards more optimal topology. Future research could focus on comparing the realized systemic risk reductions from these incentives schemes with the theoretically obtainable systemic risk minimum that is obtainable with the proposed method, and if the two methods yield similar network topologies. As we mentioned, it might be desirable to introduce additional constraints to the optimization scheme that enforce other desired economic constraints on the agents. Moreover, to ensure appropriate diversification of the agents, desired link densities can be controlled by (e.g. $L_1$-norm) constraints, or by adding penalty terms for sparsity in the objective function. 

We showed that networks restructure significantly due to the optimization. The most prominent feature is that the minimized (and maximized) networks are sparse with average link densities of around 4.4\% (11\% for the maximized), in contrast to the empirical networks that show about 56\%. Thus, we find that sparse networks can have potentially both, a very low, and a high DebtRank. Similar levels of link densities for optimal interbank networks have also been found by \cite{aldasoro2017bank} and \cite{krause2019}. 

Conversely, this means that for non-sparse systems a reduction in connectivity could result in either higher or lower systemic risk. This phenomenon should be further clarified. Non-monotonic behaviour of systemic risk as a function of link density has been observed in the literature \cite[]{nier2007network, gai2010contagion, glasserman2016contagion}. 

Another related topic in the literature on contagion is that the relationship between link density and systemic risk is usually  associated with the trade off between the diversification of risk on the individual bank level and system-wide stability. Diversification effects are usually assumed to be larger for higher link densities, which in turn can lead to higher overall systemic risk compared to networks with less risk sharing (lower link density), compare \cite{allen2000financial}, \cite{battiston2012liaisons}, \cite{battiston2012default}, or \cite{aldasoro2017bank}. Since we find that systemic risk can be high or low for a given value of the link density, our findings suggest -- consistent with intuition -- that the diversification effect on the system wide stability strongly depends on the details of how the risk is shared among banks. Future research could focus on the dependence of topological network characteristics and their relation to DebtRank for optimized and partially optimized networks. 

Let us finally mention limitations of the approach. Due to the anonymous nature of data it was not possible to estimate realistic probabilities of default in the Austrian banking system. When we assign a default probability to every bank, we do it in a static way before optimization, and assume that the rewiring of the links does not change the default probability. However, this might be not entirely unrealistic, since for every bank we keep the credit risk 
weighted exposure constant. Consequently, the credit quality of banks should not be affected by the rewiring and thus not affect the credit quality of their creditors. We mention ways to determine default probabilities more realistically  in \ref{creditrisky}. Once default probabilities are available they can be implemented in the optimization constraints as shown. Another shortcoming is that for simplicity we assumed a simple maturity scheme for financial assets. Obviously, it would be of practical interest to generalize the approach to a more realistic maturity scheme that could become valuable as soon maturities become available in interbank exposure data. Finally, we considered only a single layer of exposures. However, it is known that systemic risk may strongly depend on different layers of financial exposures  \cite[]{leon2014financial, poledna2015the, molina2015multiplex, poledna2018quantification}. An interesting extension of this present work would be to understand the effect of multiple exposure layers and their interactions on the minimum systemic risk. It is conceivable that optimization becomes technically much more challenging for multilayer networks.

\section*{Acknowledgements}
CD acknowledges funding from the WWTF project \textit{Stochastic Filtering and Corporate and Sovereign Credit Risk} project number MA14-031, PI R{\"u}diger Frey, and the OeNB anniversary fund project \textit{Dynamic measures of systemic risk}, project number 17793, PI Birgit Rudloff. ST acknowledges support from the OeNB anniversary fund project \textit{Data-driven multi-layer network approaches to quantify the spreading of systemic risk}, project number 17795,  and FFG project under 857136.

\section*{References}
\bibliography{litbib}

\newpage

\appendix

\section{Details on the MILP} \label{app:example}

\begin{figure}[t]
	\centering
	\includegraphics[scale=.45, keepaspectratio]{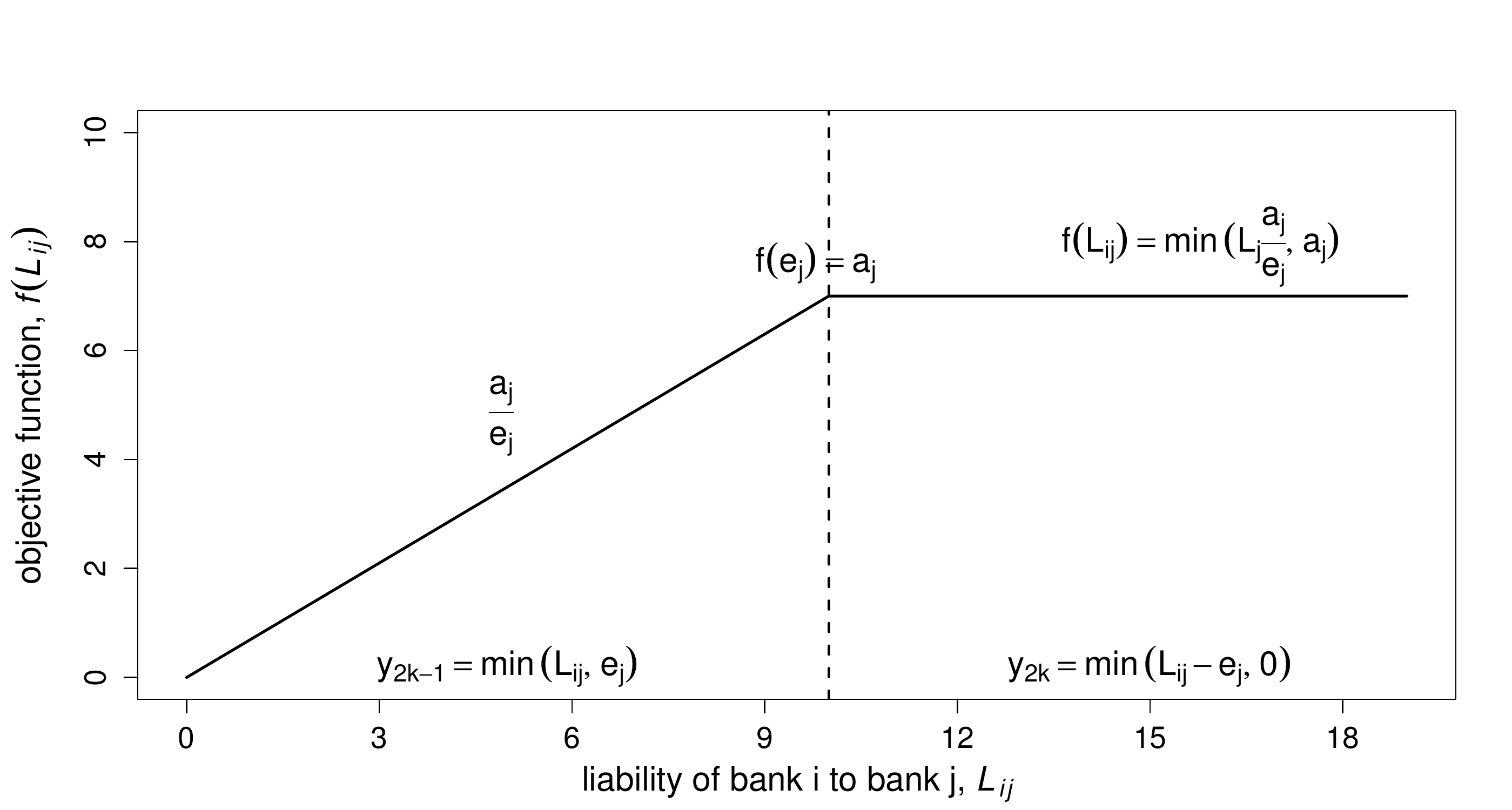}
	\caption{Visualisation of a single term of the objective function, $f(L_{ij})=\min\big(L_{ij} \frac{ a_j}{e_j}, a_j \big)$. 
		For the example we set $e_j=10, a_j=7$. The slope for $L_{ij}\in[0,10)$ is $a_j/e_j$, and 0, for $L_{ij}>10$. 
		For $L_{ij} = e_j$, we have that $f(L_{ij})=a_j$, which is the maximal functional value. 
		The two variables $y_{2k-1 }$ and $y_{2k}$ are defined as $y_{2k-1} = \min(L_{ij},e_j)$, and $y_{2k}= \min(L_{ij}-e_j,0)$. }
	\label{objectivefunction}       	
\end{figure}	
	
This section contains the details on the structure of the constraint matrices introduced in Section \ref{method} and visualizes the behavior of the objective function as mentioned in Section \ref{method}. Matrix $A_1$ contains the constraints corresponding to the reformulation of the $\min(.)$ in the objective function, Eq. \ref{eq:optim_simple}. Consider the constraints from Eq. (\ref{eq:con_delta1}) to (\ref{eq:con_delta4})
\begin{eqnarray}
y_{2i-1} & \geq & \delta_{2i} \bar{e}_i \notag  \quad, \\
y_{2i-1} & \leq & \delta_{2i-1} \bar{e}_i \notag \quad,  \\
 y_{2i}   & \leq &   \delta_{2i} u_i  \notag \quad, \\
\delta_{2i-1} & \geq & \delta_{2i} \quad, \notag 
\end{eqnarray}
where $u_i = \max \left(0, \min \left( \bar{a}_i,\bar{l}_i \right)- \bar{e}_i \right)$. We have to consider that the size of every pair $(y_{2i-1},y_{2i})$, representing one value in the liability matrix is constrained by the respective row and column sums of the matrix, and the equity value of the borrowing bank.  We recall that the correspondence of the variables is $L,x,y$ is
\begin{equation}
L_{kl} = L_{\big(i-(\ceil*{\frac{i}{n}}-1)n , \ceil*{\frac{i}{n}} \big)} = x_i = y_{2i-1} + y_{2i} \notag.
\end{equation}
	
Let's denote the upper bound for $y_{2i}$ by $u_i = \max(0, \min( \bar{a}_i,\bar{l}_i )- \bar{e}_i )$, and recall that the upper bound for $y_{2i-1}$ is $\bar{e}_i$. The first $2N^2$ columns correspond to the $y$ variables. The second $2N^2$ columns correspond to the vector of dummy variables $\delta$, which indicate if the respective $y_j$ is positive or zero. For each entry of the matrix $L$ and the corresponding $x_i$ there are four rows enforcing the constraints from Eq. (\ref{eq:con_delta1}) to (\ref{eq:con_delta4}). Note that the diagonal elements of $L$ are forced to zero by specifying the upper bound and lower bound in the optimization software to zero. The structure of $A_1$ is schematically depicted as
{\tiny
\begin{equation*}
A_1=
\begin{array}{c|cccccccccccccccc}
& & & & & & & & & & & & & & & & \\ 
& 1 & 2 & ... &  2i -1 &  2i  & ... &  2N^2 -1 &  2N^2  &  2N^2 +1 &  2N^2 +2 & ... &  2N^2 +2i-1 &  2N^2 +2i & ... &  4N^2 -1 &  4N^2  \\ \hline
1 & 1 & 0 & 0 & 0 & 0 & 0 & 0 & 0 &  -\bar{e}_1  & 0 & 0 & 0 & 0 & 0 & 0 & 0 \\ 
2 & -1 & 0 & 0 & 0 & 0 & 0 & 0 & 0 & 0 &  \bar{e}_1  & 0 & 0 & 0 & 0 & 0 & 0 \\ 
3 & 0 & 1 & 0 & 0 & 0 & 0 & 0 & 0 & 0 &  -ub_1  & 0 & 0 & 0 & 0 & 0 & 0 \\ 
4 & 0 & 0 & 0 & 0 & 0 & 0 & 0 & 0 & -1 & 1 & 0 & 0 & 0 & 0 & 0 & 0 \\ 
... &  &  &  &  &  &  &  &  &  &  &  &  &  &  &  &  \\ 
4i-3 & 0 & 0 & 0 & 1 & 0 & 0 & 0 & 0 & 0 & 0 & 0 &  -\bar{e}_i  & 0 & 0 & 0 & 0 \\ 
4i-2 & 0 & 0 & 0 & -1 & 0 & 0 & 0 & 0 & 0 & 0 & 0 & 0 &  \bar{e}_i  & 0 & 0 & 0 \\ 
4i-1 & 0 & 0 & 0 & 0 & 1 & 0 & 0 & 0 & 0 & 0 & 0 & 0 &  -ub_i  & 0 & 0 & 0 \\ 
4i & 0 & 0 & 0 & 0 & 0 & 0 & 0 & 0 & 0 & 0 & 0 & -1 & 1 & 0 & 0 & 0 \\ 
... &  &  &  &  &  &  &  &  &  &  &  &  &  &  &  &  \\ 
4N^2 -3 & 0 & 0 & 0 & 0 & 0 & 0 & 1 & 0 & 0 & 0 & 0 & 0 & 0 & 0 &  -\bar{e}_{N^2}  & 0 \\ 
4N^2 -2 & 0 & 0 & 0 & 0 & 0 & 0 & -1 & 0 & 0 & 0 & 0 & 0 & 0 & 0 & 0 &  \bar{e}_{N^2}  \\ 
4N^2 -1 & 0 & 0 & 0 & 0 & 0 & 0 & 0 & 1 & 0 & 0 & 0 & 0 & 0 & 0 & 0 &  -ub_{N^2}  \\ 
4N^2  & 0 & 0 & 0 & 0 & 0 & 0 & 0 & 0 & 0 & 0 & 0 & 0 & 0 & 0 & -1 & 1. \\
& & & & & & & & & & & & & & & & \\
\end{array} 
\end{equation*} }	
\vspace{1cm}

\noindent The matrix $A_2$ is responsible for the column sum constraint $A_2 z = a$. Note that $A_2 \in \mathbb{R}_+^{N\times 4N^2}$, where the last $2N^2$ columns are zero columns as the binary vector is not in use here. Since, the diagonal elements of $L$ are set to zero the correct column sum must be achieved through the other $N-1$ entries in the respective column. Since, $x = (L_{11},\dots,L_{N1}, L_{12}\dots,  L_{N2}, \dots,  L_{1N},\dots,L_{NN})^\top$, the first $N$ entries of $x$ correspond to the first column of $L$, the next $N+1$ to $2N$ entries to the second column and so on. This translates again to $y$ and the first $2N^2$ columns of $A_2$ follow the diagonal structure below. For a more compact notation\footnote{In the following we slightly abuse notation and use the subscript for defining the dimension of vectors having a constant value.} of $A_2$, we define the vectors $\iota_i$ of length $2N$ as	
\begin{eqnarray}
	\iota_1 & = & (0_2,1_{2N-2}) \notag \\
	\iota_2 & = & (1_2,0_2,1_{2N-4}) \notag \\
	\iota_i & = & (1_{2i-2}, 0_2, 1_{2N-2i}) \notag \\
	\iota_n & = & (1_{2N-2},0_{2}) \notag .
\end{eqnarray}
Where $0_2 = (0,0)$ and $1_{i} = (1,\dots, 1)$ of length $i$. Then, $A_2$ has the following structure 	
\begin{equation}
	 A_2 = 
	 \begin{bmatrix}
	 \iota_1 & 0 & \dots & 0 & 0 & 0 & 0_{2n^2} \\ 
	 0 & \iota_2 & 0 & \dots & 0  & 0 & 0_{2n^2} \\ 
	 \vdots & 0 & \ddots & \ddots &  & \vdots & 0_{2N^2} \\ \notag
	 0 & \vdots & \ddots & \iota_i & 0 & 0 & 0_{2N^2} \\ 
	 0 & 0 &  & 0 & \ddots & 0 & 0_{2N^2} \\ 
	 0 & 0 & \dots & 0 & 0 & \iota_N & 0_{2N^2}
	 \end{bmatrix},
\end{equation}

\vspace{3mm}
\noindent
and since the rows of $A_2$ are of length $4N^2$, $0$ must be of length $2N$, as $\iota$. 
 
Matrix $A_3$ enforces the constraints on row sums and has a similar structure as $A_2$. Again the first $2N^2$ entries correspond to the $y$ values and the latter ones are corresponding to the binary variables and are zero. Again, we set $0_2 = (0,0)$ and $1_2=(1,1)$, and  define a sequence of auxiliary matrices $B_1, B_2 \dots, B_N$ with dimension $N\times 2N$ as
    
\begin{equation}
    B_1 = 
    \begin{bmatrix}
    0_2 & 0_2 & \dots & 0_2 & 0_2 & 0_2 \\ 
    0_2 & 1_2 & 0_2 & \dots & 0_2 & 0_2 \\ 
    \vdots & 0_2 & \ddots & \ddots &  & \vdots \\ 
    0_2 & \vdots  & \ddots & 1_2 & 0_2 & 0_2 \\ 
    0_2 & 0_2&  & 0_2 & \ddots & 0_2 \\ 
    0_2 & 0_2 & \dots & 0_2 & 0_2 & 1_2
    \end{bmatrix}  \notag.
\end{equation}
    
\begin{equation}
    B_2 = 
    \begin{bmatrix}
    1_2 & 0_2 & \dots & 0_2 & 0_2 & 0_2 \\ 
    0_2 & 0_2 & 0_2 & \dots & 0_2 & 0_2 \\ 
    \vdots & 0_2 & 1_2 & \ddots &  & \vdots \\ 
    0_2 & \vdots  & \ddots & \ddots & 0_2 & 0_2 \\ 
    0_2 & 0_2&  & 0_2 &1_2 & 0_2 \\ 
    0_2 & 0_2 & \dots & 0_2 & 0_2 & 1_2
    \end{bmatrix}  \notag.
\end{equation}
    
\begin{equation}
     B_i = 
     \begin{bmatrix}
     1_2 & 0_2 & \dots & 0_2 & 0_2 & 0_2 & 0_2 \\ 
     0_2 & \ddots & 0_2 & \dots & 0_2 & 0_2 & 0_2\\ 
     \vdots & 0_2 & 1_2 & \ddots &  & \vdots & 0_2\\ 
     0_2 & \vdots  & \ddots & 0_2 & 0_2 & 0_2 & 0_2\\ 
     0_2 & 0_2&  & 0_2 & 1_2 & \ddots & 0_2\\ 
     0_2 & 0_2 & \dots & 0_2 & \ddots & \ddots & 0_2\\
     0_2 & 0_2 & 0_2 & 0_2 & 0_2 & 0_2 & 1_2
     \end{bmatrix}  \notag.
\end{equation}
    
Then $A_3$ can be defined as block matrix
\begin{equation}
    A_3 =
    \begin{bmatrix}
    B_1 & B_2 & \dots & B_i & \dots & B_N & 0_{2N^2 \times 2N^2}.\\
    \end{bmatrix}
\end{equation}
Matrix $A_4$ is responsible for keeping the credit risk weighted exposure to other banks constant for each interbank loan portfolio. Recall that the credit riskiness of bank $i$ is $\kappa_i$, and the credit risk weighted exposure of the empirically observed matrix is $r=L^\top \kappa$. Then, $A_4$ is a $\kappa$-weighted version of $A_2$. Let 
\begin{eqnarray}
\nu_1 & = & \iota_1 \cdot \kappa \notag \\
\nu_2 & = & \iota_2 \cdot \kappa \notag \\
	\vdots & \vdots & \quad \vdots \notag \\
	\nu_N & = & \iota_N \cdot \kappa \notag  ,
\end{eqnarray}
where $\cdot$ denotes the pointwise multiplication of two vectors. Then $A_4$ can be defined as 	
\begin{equation}
	 A_4 = 
	 \begin{bmatrix}
	 \nu_1 & 0 & \dots & 0 & 0 & 0 & 0_{2N^2} \\ 
	 0 & \nu_2 & 0 & \dots & 0  & 0 & 0_{2N^2} \\ 
	 \vdots & 0 & \ddots & \ddots &  & \vdots & 0_{2N^2} \\ 
	 0 & \vdots & \ddots & \nu_i & 0 & 0 & 0_{2N^2} \\ 
	 0 & 0 &  & 0 & \ddots & 0 & 0_{2N^2} \\ 
	 0 & 0 & \dots & 0 & 0 & \nu_n & 0_{2N^2}
	 \end{bmatrix}  .
\end{equation}
This completes the set of constraints for the MILP.

\section{Network visualization and topology measures} \label{topology}

This section contains the details on the network visualization and provides additional figures on the effects of the optimization on the network structure. 	
\begin{figure}[t]	
\centering
	\includegraphics[scale=.4, keepaspectratio]{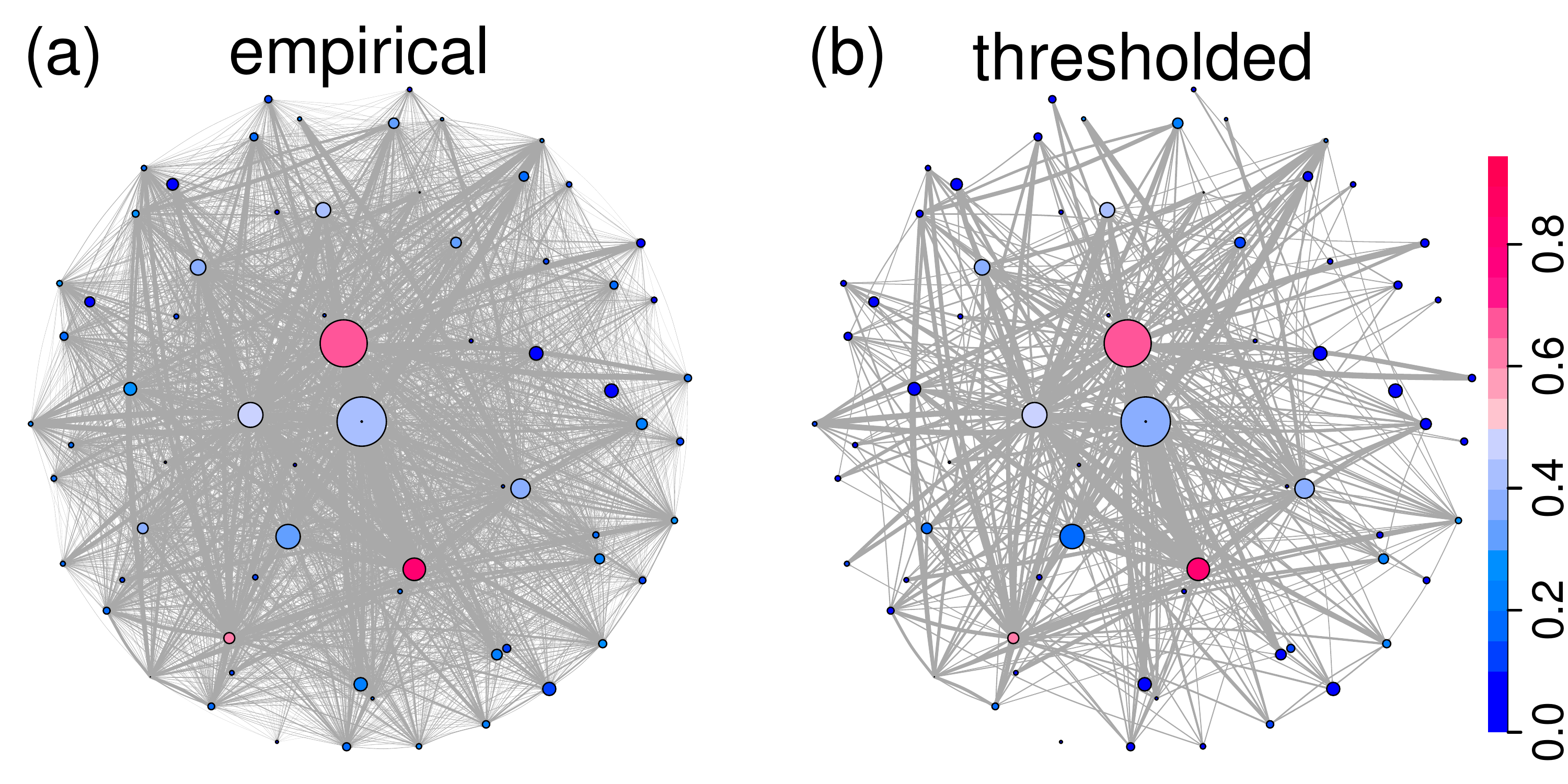}
	\includegraphics[scale=.4, keepaspectratio]{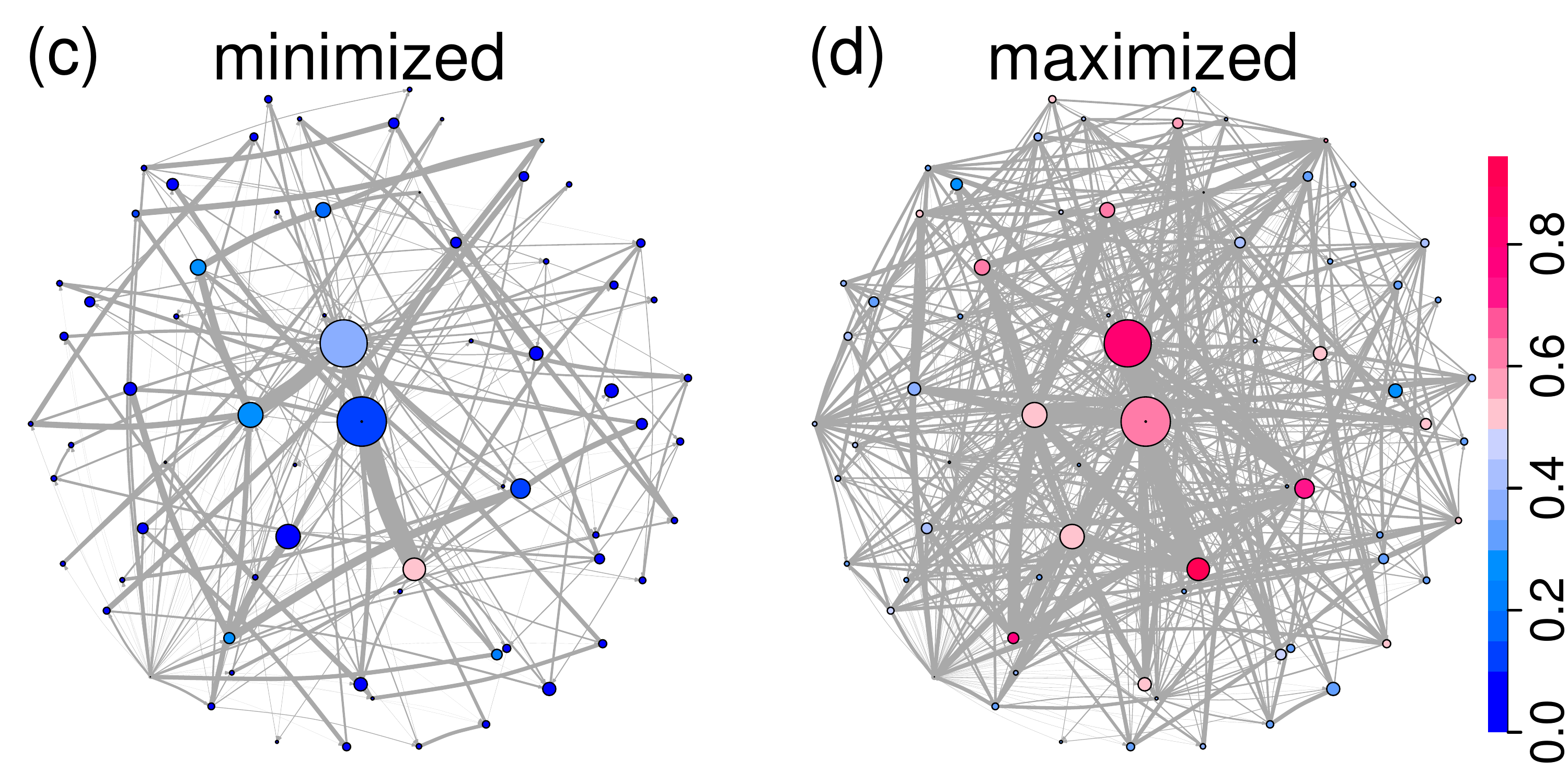}
	\caption{Comparison of the empirical asset-liability network, $L$, as observed in Q1 (a), with a thresholded network, where 
	 the 454 largest links are shown that cover 90\% of the interbank market volume. 
	 The link density decreases from around 62\% in (a) to 9.4 \% in (b), which is visibly sparser.
	  (c) Minimized network (same as in Figure \ref{densityassortativity} (b)). 
	  (d) Maximized network in the same quarter. 
	The color represents a high DebtRank (red), low systemic risk is blue. 
	  }
\label{empnwplot}		
\end{figure}

Figure \ref{empnwplot} (a) shows a visualization of the empirically observed network of the 70 largest Austrian banks for quarter Q1, same as Figure \ref{densityassortativity} (a) in the main text. The link density is around 62\%. (b) shows the same network, but thresholded such that only the largest links accounting for 90\% of interbank market volumes are used. This reduced network has an link density of about 9.4\%. Figure \ref{empnwplot} (c) shows the minimized network for the same quarter Q1, which has a link density of 4.4\%. The respective maximized network is visualized in Figure \ref{empnwplot} (d). It has an link density of around 12\%. In all four panels node size corresponds to the equity of the node, and the link width corresponds to the size of the respective liability, $L_{ij}$. Since link- and  equity values differ strongly in size, we employed transformations of the quantities to provide a more readable presentation of the networks. Values are first compressed by taking the square root  and then standardized by the maximal edge size, $ L_{ij}^{\rm max} = \max \{L_{ij}^{0.5} | \forall ij \in \{1,\dots,n\}^2 \}$, and maximal equity size, $E_i^{\rm max} = \max \{E_{i}^{0.5} | \forall i \in \{1,\dots,n\} \} $, to get values between zero and one. Then we multiply by a factor of 15. Dark red and dark blue nodes indicate high and low DebtRank $R_i$, respectively, lighter tones show medium sized values. Network visualization were produced with the igraph R package \cite{csardi2006igraph}.

\begin{figure}[h]
	\vspace{-0cm}
	\centering
	\includegraphics[scale=.45, keepaspectratio]{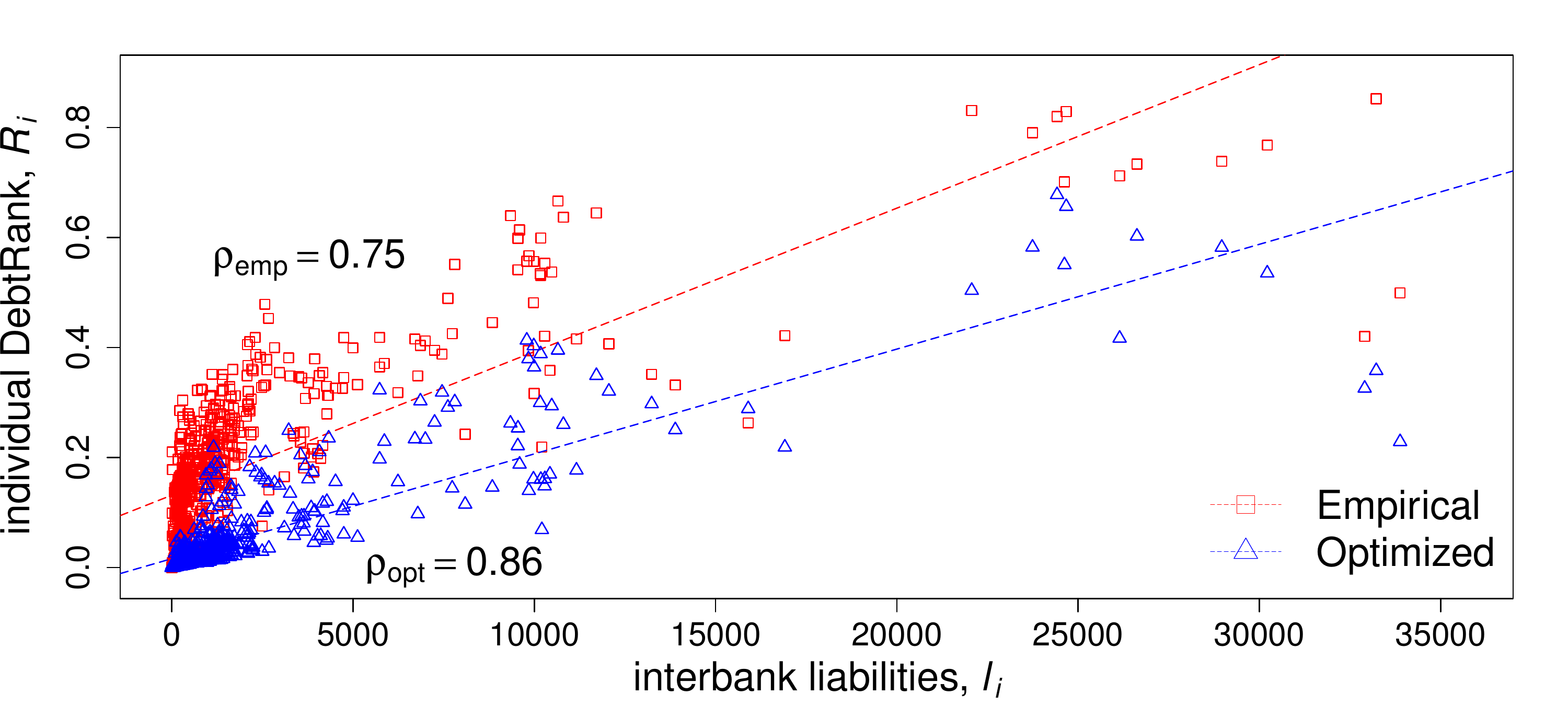}
	\caption{Comparison of individual bank DebtRanks, $R_i$, with interbank liabilities $l_i$ for the 
	empirical (red squares) and the minimized (blue triangles) 
	networks for all ten quarters. It is clear that the optimization strengthens the relationship between banks' interbank liabilities 
	$l_i$ and their DebtRank $R_i$. In saver networks systemic risk is more proportional to bank size, than in the riskier empirical ones. 	
	The optimization shows smaller banks can be rendered systemically negligible by changing the network topology.  		
	}
	\label{drsizecomparison} 
\end{figure}

Figure \ref{drsizecomparison} plots the individual bank DebtRanks $R_i$ against the banks' interbank liabilities $l_i$ for all quarters. In general DebtRanks $R_i$ are higher for banks with large interbank liabilities $l_i$. It is clearly visible that in the optimized networks bank DebtRanks are more proportional to the banks' interbank liabilties than in the original network. This is also supported by the higher correlation of 0.86 for the optimized network than for empirical network where the correlation is 0.75. For the empirical network the relationship looks non linear and small banks ``punch highly above their weight". The optimization resolves this issue and renders small banks systemically negligible. 

\begin{figure}[t]	 	
\centering
	\includegraphics[scale=.28, keepaspectratio]{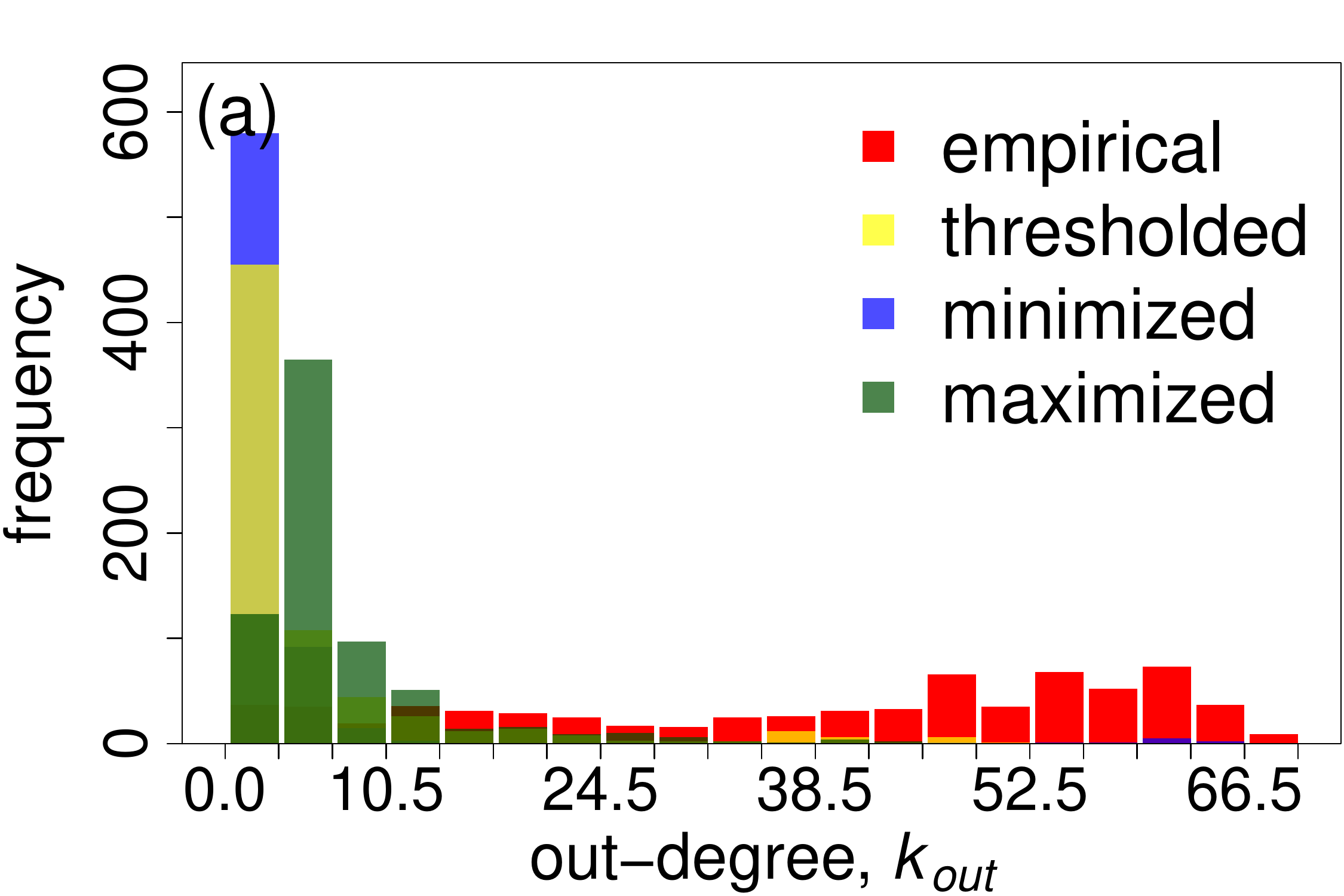}
	\includegraphics[scale=.28, keepaspectratio]{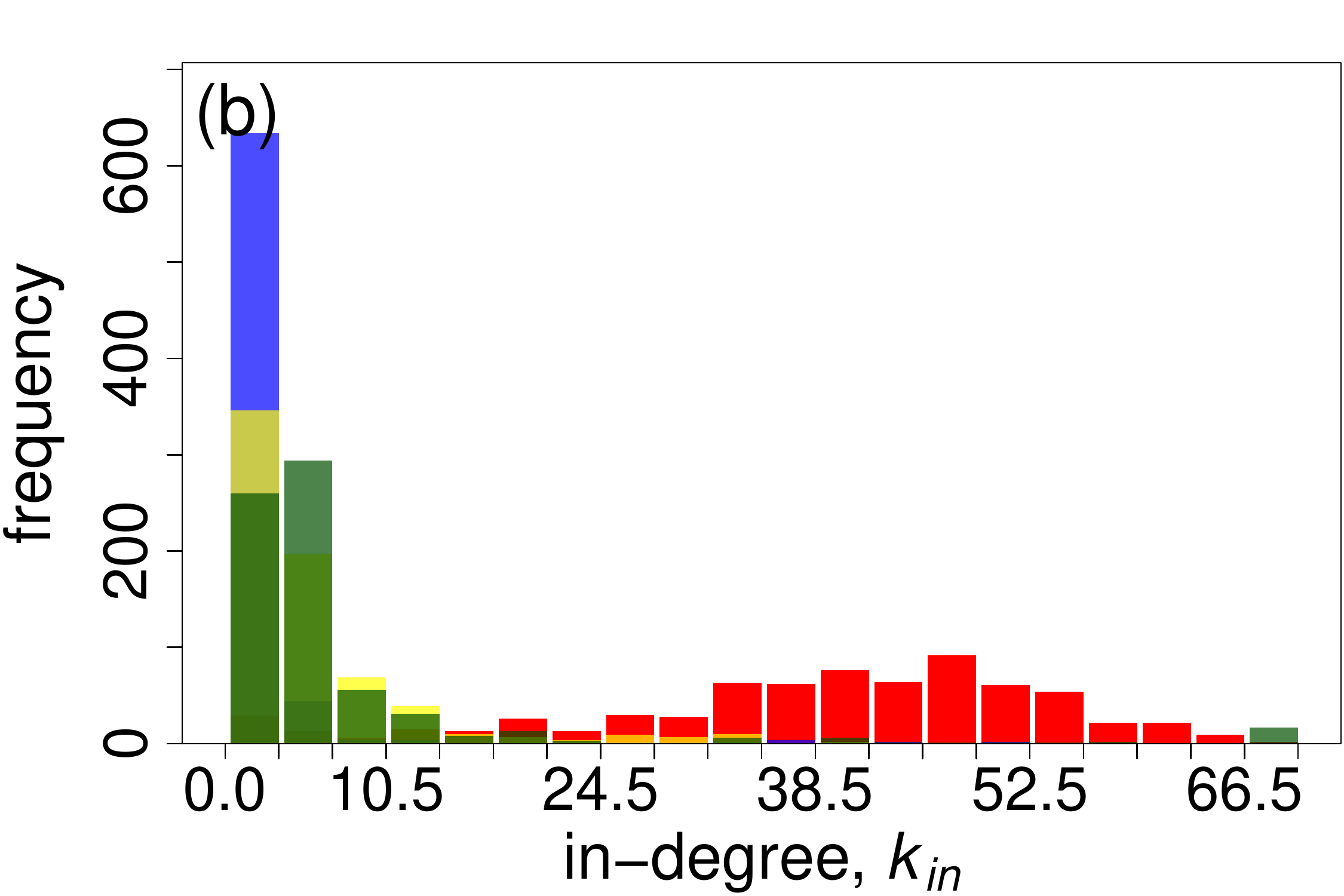}	\\
	 \includegraphics[scale=.28, keepaspectratio]{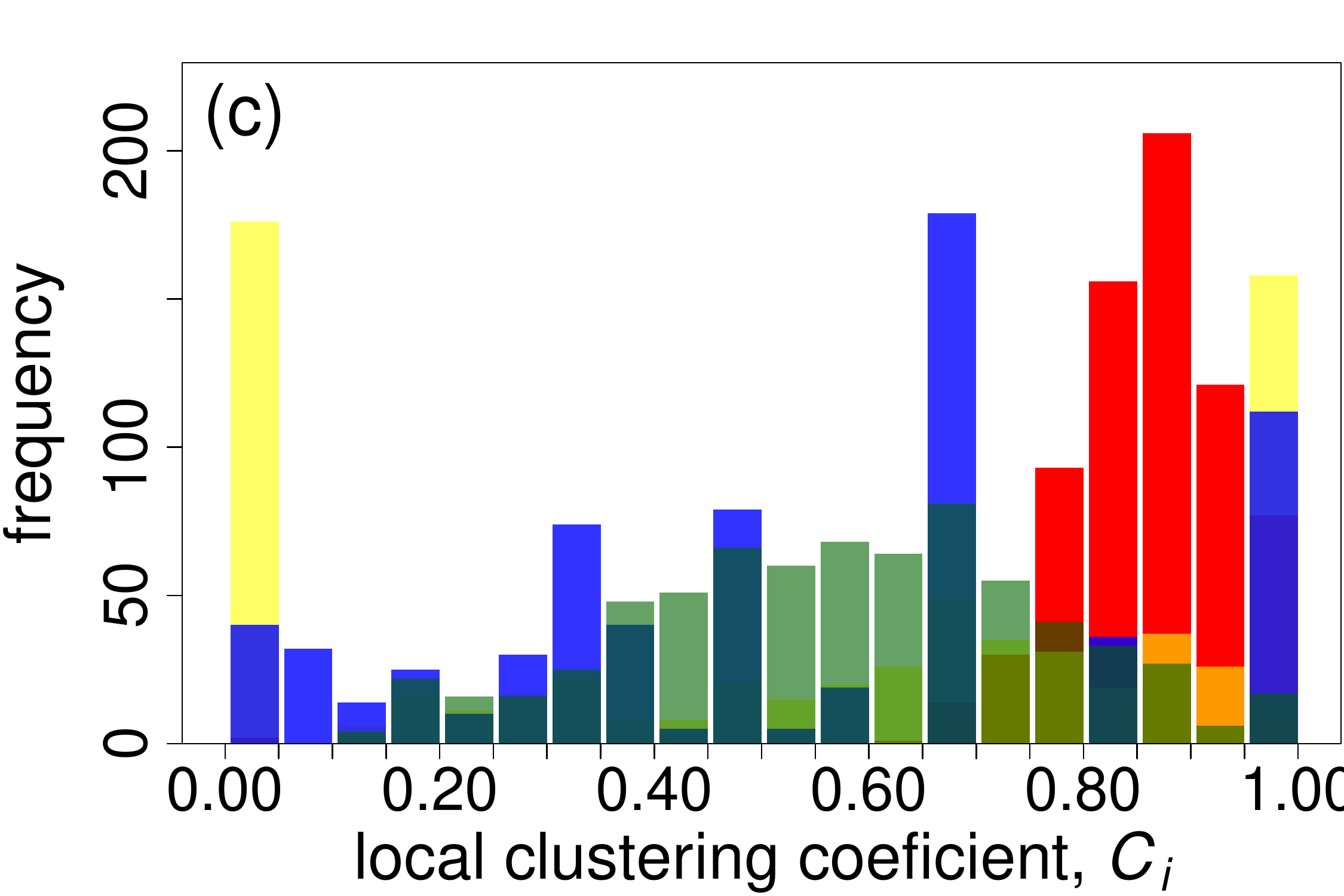}
	 \includegraphics[scale=.28, keepaspectratio]{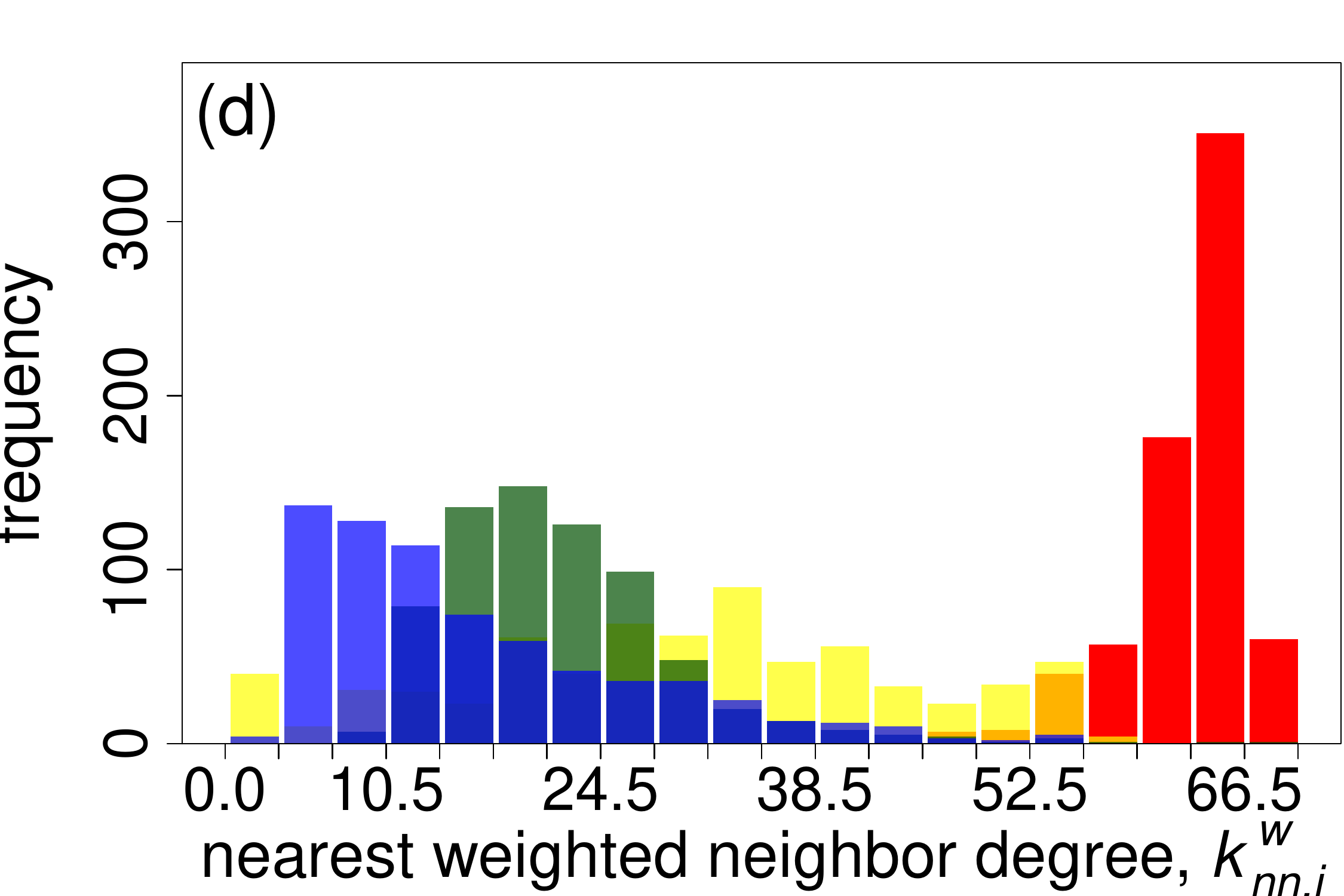} 
	\caption{ (a) shows the distribution of the out-degrees, $k_{\rm out}$, for the four different 
	network types, empirical (red), minimized (blue), maximized (green), thresholded (yellow). 
	Data is pooled for the ten quarters. 
	(b) in-degree distribution of $k_{\rm in}$ for the same networks
	(c) Distribution of the local clustering coefficient, $C_i$. There are obvious differences. 
	(d) Distribution of the nearest weighted neighbor degree, $k_{nn,i}^w$.
	}
	\label{inoutdeg_dist}
\end{figure}

The following Figure \ref{inoutdeg_dist} provides a more detailed perspective on the local network properties such as the in- and our- degree distributions the clustering behavior, and the weighted nearest neighbor degree for the four different network types. The measures are pooled over all ten observations. Figure  \ref{inoutdeg_dist} (a) shows the distribution of the out-degrees $k_{out}$ for the four different network types across all ten observations; empirical (red), minimized (blue), maximized (green), thresholded (yellow). (b) shows the distribution of the in-degrees $k_{in}$ for the four different network types across all ten observations. In both cases, degrees of the minimized networks are severely  peaked in the range of 0 and 3.5. Only at the value of 70 there is small mass for the out-degree. The maximized networks peak between 3.5 and 7 and are right-skewed. There is small mass at 70 for the in-degree. The empirical network degrees have a wide distribution with most mass around 55 for the out-degree and 45 for the in-degree. The thresholded network degrees are peaked between 0 and 3.5 and decay fast.
Figure \ref{inoutdeg_dist}  (c) shows the distribution of the local clustering coefficient $C_i$ for the four different network types. There are obvious differences. The empirical clustering coefficient distribution has the bulk of its  mass at values around 0.85, whereas the thresholded network is bi modal and has its mass equally divided between zero and one. The distribution for the minimized networks has its mass distributed across the whole spectrum with peaks at 0.7 and 1 and the remainder of the mass is allocated at smaller values. The distribution of the maximized networks is rather flat and distributed between 0.2 and 0.8 with a peak around 0.55. (d) shows the distribution of the weighted nearest  neighbor degree $k_{nn,i}^w$ for the same networks. Again,  there are obvious differences. The $k_{nn,i}^w$ for the empirical networks are centered around 65, whereas the thresholded network values are distributed between 0 and 60 with a slight peak at around 30. The values of the minimized network are right skewed with a peak around 7. The values for the maximized networks are ranging between 10 and 50 with a peak at around 20.

\section{Comparing results with DebtRank2 \label{DR2}}
	
\begin{figure}[t]				
	\centering
	\includegraphics[scale=.45, keepaspectratio]{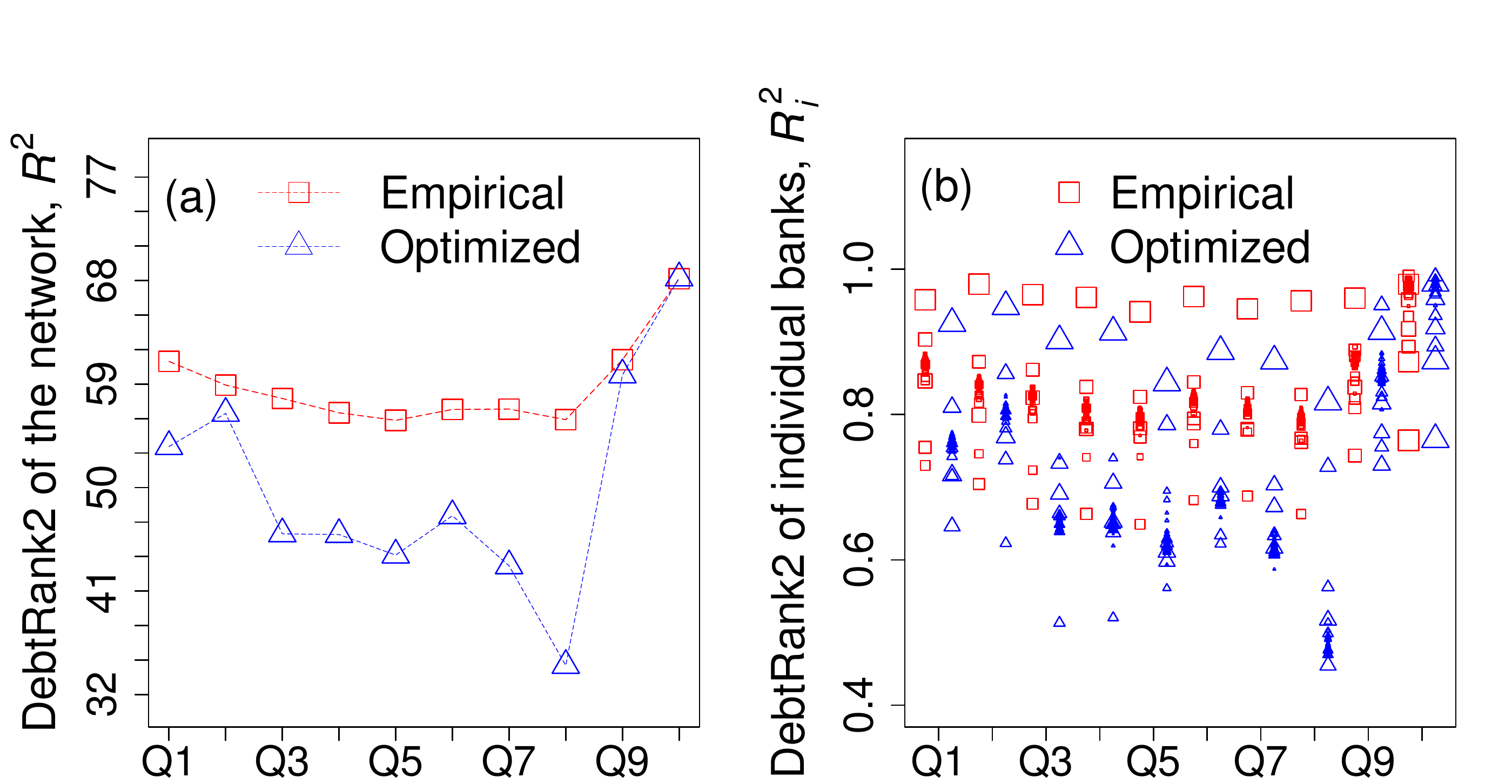}
	\caption{Same as Figure \ref{drempvsmin} (a) and (b) but for the definition of DebtRank2. 
	In comparison to the ordinary DebtRank we see that the DebtRank2 levels are 
	substantially higher for both the emprical and the minimized networks. In quarter 
	Q10 both the empirical and the minimized DebtRank apporach the theoretical maximum of 
	DebtRank 69 for a network of 70 banks. The minization works best for Q8 where the ratio of 
	interbank market volume and equity $\bar{L}/\bar{E}$ is smallest. 
	(b) shows that the optimization works better for smaller banks. 
	}
	\label{dr2empvsmin} 
\end{figure}

This section investigates the robustness of the proposed optimization with respect to the definition of  DebtRank. For this reason we substitute the original DebtRank defined in \cite{battiston2012debtrank} by a variation that was proposed in \cite{bardoscia2015debtrank}, which we call DebtRank2. We find that the minimization procedure is still producing networks with lower systemic risk when DebtRank is interchanged with DebtRank2.  However, there are differences. DebtRank2 $R^2$ is reduced on average by 15\% from 59 to 50. We present the corresponding plot to Figure \ref{drempvsmin} (a) and (b) in Figure \ref{dr2empvsmin}. The first observation is that the levels for DebtRank2 are substantially higher than for the original DebtRank.  Additionally we see that when the ratio $\bar{L}/\bar{E}$ increases, as shown in Figure \ref{drempvsmin} (c), the systemic risk reduction potential for DebtRank2 is diminishing in the last two observations. Furthermore, in \ref{dr2empvsmin} (b) we can observe that the single bank DebtRanks are more homogenous in comparison to Figure \ref{drempvsmin} (b). 
			
\section{Estimating credit riskiness and data needs} \label{creditrisky}

Since in the empirical part of this paper we focus on a direct exposure network, the specification of the credit risk variable, $\kappa$, needs to be considered. We suggest three state-of-the-art approaches to specify the credit riskiness of individual banks, which are found in \cite{mcneil2015quantitative}. However, all of them have specific data needs, and to a large extent, this data is usually not publicly available. Thus we need to employ a proxy for $\kappa$ instead. The first and probably most broadly applicable choice for $\kappa$ are ratings, which are assigned to banks by rating agencies like Standard and Poors, Moodies and Fitch, which consider historical probabilities of default (PDs) for the respective rating class as $\kappa$. The disadvantage of this is that agencies usually update their ratings gradually and ratings are sticky to a certain extent. In suddenly deteriorating economic environments of financial crises, the ratings might not reflect the actual risk. 
The second approach suggested is using an analytical model, which can be calibrated to publicly available data. One of the most popular structural credit risk models is the Merton model \cite[]{merton1974pricing}. Several versions are used in practice to calculate PDs. The advantage of the model is that it can be calibrated with publicly available balance sheet data and stock market prices. The incorporation of stock market data captures most recent information. The third approach would be to consider CDS spreads of banks and infer the PDs from a reduced form credit risk model, which is calibrated to the observed CDS spreads. In the reduced form approach the CDS spreads used to calibrate the model take into account current market information. 

All of these approaches are generally difficult to put into practice, since for small banks there might be insufficient information available to calibrate an analytical model. A possible solution could be to use credit risk estimates of banks, where the data is available and perform a regression  analysis with accounting ratios as independent variables. The so fitted model could be used to obtain an estimate for the banks for which the necessary information is not accessible.

\section{Equality vs. inequality in the credit risk constraint} \label{equivalence}

\begin{align} 
\min_{L \;\in \; \{M \;\in \; \mathbb{R}_+^{N \times N} \; : \; diag(L)=0 \} } &  \sum_{i=1}^{N}\sum_{j=1}^{N} \min \left( \frac{L_{ij}}{e_j},1 \right) \frac{a_j}{\bar{L}} \notag  \tag{P1} \\
\text{subject to:} \qquad & l_i = \sum_{j=1}^{n} L_{ij} \quad \forall i \tag{C1} \quad, \\ 
&  a_i = \sum_{j=1}^{N} L_{ji} \quad \forall i \nonumber \tag{C2} \quad,\\
&  r_i = \sum_{j=1}^{N} L_{ji}\kappa_j \tag{C3} \quad.
\end{align}

We show that in the formulation of the optimization problem (P1) changing  the credit risk constraint (C3) from equality  ($=$) to greater or equal than ($\geq$), does not change the solution, when the row and column sum constraints (C1) and (C2) are in place. This is intuitively clear, because the row sum constraint keeps the amount of interbank liabilities in the system constant. This implies that also the total sum of credit risk weighted liabilities $\sum_{j=1}^{N}r_j = \sum_{j=1}^{N}\kappa_j l_j $ must remain constant. If now one or more banks would have a lower credit risk weighted exposure $r_i$ after the optimization than before the optimization and none of the banks has a higher risk weighted exposure $r_j$ (as implied by the $\geq$ constraint) this means that the total risk weithed exposure $\sum_{j=1}^{N}r_j$ must be smaller after the optimization than before the optimization, which is a contradiction to the statement that $\sum_{j=1}^{N}r_j$ must remain constant, because the $l_i$ remain constant.

\end{document}